# Reasoning effort, not tool access, buys first-try reliability in agentic code generation: Evidence from 90 matched agent runs

Achint Mehta

Independent Researcher, Austin, TX, USA

Corresponding author. E-mail: achintmehta@gmail.com

## Abstract

Agentic coding assistants are increasingly given extra capabilities, such as browser-based testing tools and design-oriented system prompts, on the assumption that more capability yields better software. This study tested that assumption directly. Ninety independent agent runs built the same application, a real time retrospective board, from one detailed specification written for this study, and each was scored on a fixed 14 criterion functional rubric (42 point maximum) together with a visual quality review. The runs spanned several model generations, two agent harnesses, two reasoning effort levels in the matched contrasts, a testing tool, and two design-oriented prompts, with identical configurations repeated and one factor varied at a time. Conditions were chosen rather than randomized. Capability tier dominated outcomes: frontier models clustered near the ceiling while two low-cost local models scored 37 and 24 points. A criterion level analysis added results that run totals conceal. Container deployment was the dominant defect, failing on the first attempt in 44 percent of runs, and its failure rate shifted sharply across model generations while mean totals moved by less than a point. The testing tool raised cost by 42 to 68 percent without improving functional score or reliability, even on interface visible criteria, with the premium going to context re-reading rather than additional code. Raising reasoning effort from High to xHigh lifted first try perfect runs from 28 percent to 89 percent (a difference of 61 percentage points, 95 percent confidence interval 29 to 78) and cut corrective prompts about five-fold, for 9 to 29 percent more cost. A design-oriented prompt changed what the agents built without changing what worked: every prompted run shipped gradients, keyframe animations and imported display fonts and defined about three times as many design tokens, no unprompted run did any of these, and a one paragraph paraphrase of the prompt's design instruction reproduced the pattern, indicating the capability needs only invocation. Functional scores were unchanged while first try reliability fell. A model-generated visual rating agreed with this source-based classification in every Claude Code run. The practical lesson is to match the fix to the failure: most first run failures came from weak reasoning, which a stronger model or more effort prevents, not from visible flaws a checking tool would catch. The findings describe delivery reliability on this one task, mostly on one vendor's models, and they complement cross task benchmarks rather than replace them.

## Introduction

Teams that build software with agentic coding assistants often add capabilities to the agent in the hope of better results. Two common additions are a browser-based tool that lets the agent drive and inspect the running interface, and a system prompt that pushes the agent toward polished visual design. The implicit assumption is that each addition pays off in the way it is meant to, the testing tool by making the software more reliable and the design prompt by making the interface more polished, and that the gain is worth its added cost. Testing that assumption on a like for like basis is costly, because it requires running the same task many times under controlled conditions.

Public comparisons of coding agents often use a single short instruction and report a single run per model. That design hides two things that matter in practice. It hides the run-to-run variation that a single sample cannot show, and that variation is known to be large for code generation even at settings meant to be deterministic (Ouyang et al., 2025; Song et al., 2025). It also conflates the choice of model with the choice of tools, prompt, and reasoning effort, and it usually leaves cost out of the comparison altogether (Kapoor et al., 2024). This study used a different design. Every run received the same detailed specification, identical configurations were repeated several times, and one factor was varied at a time. This separates the effect of a model from the effect of a tool, a prompt, or an effort setting, and it shows how much identical configurations differ from each other. The specification was written for this study rather than taken from a public benchmark, because models have been shown to remember public benchmark material rather than reason about it (Riddell, Ni & Cohan, 2024; Liang, Garg & Zilouchian Moghaddam, 2025).

This study makes five contributions. The first is a matched, replicated contrast for a visual automation and testing tool, holding model and effort fixed, at two capability tiers. The second is a matched contrast of High and xHigh reasoning effort on a model that exposes both settings. The third is a judge-free measurement of the design treatment in each run's shipped source, reported alongside a model-generated visual rating, which separates the effect of a design prompt on what is built from its effect on function. The fourth is a criterion level analysis of the same runs, which shows where failures occur, how failure modes shift between model generations, and why aggregate scores understate those shifts. Recent work has argued that scores which tolerate repair or many attempts hide reliability differences (Jiang et al., 2026; Khanal, Tao & Zhou, 2026), and this study adds matched, criterion level evidence of that effect from a task no model had seen. The same per run session records support a decomposition of what each added capability actually spends tokens on. The fifth is a prompt ablation, meaning a comparison that strips an intervention down to the part thought to carry its effect and re-runs it to see whether the effect survives. Here a one paragraph abridged directive written for this study is compared against the full third-party prompt whose design instruction it paraphrases, testing whether the prompt's length and emphatic tone are necessary for its effect. The stripped-down condition is a paraphrase rather than a verbatim excerpt, so it varies wording alongside length and tone. The next section places these contributions in prior work.

## Related work

## Benchmarks and the validity of their scores

Most measurement of code generation happens on standardized benchmarks. HumanEval introduced pass@k, the chance that at least one of k attempts passes the tests (Chen et al., 2021), and MBPP uses the same short function format (Austin et al., 2021). Later benchmarks grew the unit of work to whole classes, where skill at one size did not predict skill at the other (Du et al., 2023), and to tasks combining many libraries, where models fall well short of human performance (Zhuo et al., 2025). AlphaCode reached a competitive contest level only by generating and filtering many candidates, an early sign that what a model can reach with many attempts and what it delivers on one are different things (Li et al., 2022). A second generation of benchmarks made the setting agentic: instead of completing a function from a prompt, the model drives an agent that reads, edits and runs code in a repository, so what is scored is what the agent delivers rather than what the model can write in one pass. SWE-bench set the pattern by asking agents to resolve real issues drawn from GitHub projects, judged by those projects' own test suites (Jimenez et al., 2024). SWE-agent then showed that the interface through which an agent views files and runs commands changes its success rate, so results depend on the harness as well as on the model (Yang et al., 2024), a point that motivates holding the harness fixed here. OpenHands provides an open platform on which such agents can be built and compared (Wang et al., 2025), and Terminal-Bench evaluates agents on realistic tasks carried out entirely in a command line environment, the setting used in this study (Merrill et al., 2026). Design2Code extends the idea to front end work, asking models to reproduce a web page from a screenshot and scoring the result by its visual similarity to the original (Si et al., 2025). Systematic reviews map the broader use of language models in software engineering (Hou et al., 2024) and its agent branch in particular (Liu et al., 2024).

A growing body of work questions what these scores measure. Hand screening of SWE-bench found most tasks underspecified or unfairly tested, producing the smaller Verified subset (Chowdhury et al., 2024). A later audit found roughly a third of passing patches had the solution given away in the issue text (Aleithan et al., 2024). Models have been caught locating the buggy file from memory rather than from the issue (Liang, Garg & Zilouchian Moghaddam, 2025), and benchmark and training data overlap has been measured directly (Riddell, Ni & Cohan, 2024; Matton et al., 2024). Strengthening the HumanEval tests dropped pass rates by 19 to 29 percent and changed rankings (Liu et al., 2023), and such problems are widespread enough to change leaderboard order (Zhu et al., 2025), with fresh problems that postdate model training one proposed remedy (Jain et al., 2025). This study is not a benchmark and responds differently: it runs one full application specification, written for this study so there is nothing to remember, many times under matched configurations, and checks every criterion by running the delivered application. It trades breadth across tasks for repetition on one task, and measures what a configuration delivers on the first attempt rather than what a model family can reach across many.

## First-attempt reliability and reasoning effort

What a model delivers on its own and what it can reach with help are different quantities. Self-repair yields gains that are modest and inconsistent, limited by the model's ability to diagnose its own faults (Olausson et al., 2024), and models correct their own reasoning poorly without outside feedback (Huang et al., 2024). Recent work argues the usual summary numbers hide this: pass@k is often computed in ways that make agents look more reliable than they are, so reliability should be measured over independent repeated runs (Jiang et al., 2026), and evaluation of long running agents has been reframed around keeping capability and reliability separate (Khanal, Tao & Zhou, 2026). This study adds matched, criterion level evidence of the same effect under a grading scheme that allows repair: run totals less than a point apart concealed first try pass rates differing by more than fifty percentage points on the dominant defect.

The effort setting studied here follows work showing that models do better when they spend more computation before answering: reasoning step by step improves accuracy (Wei et al., 2022), sampling several paths and taking the majority improves it further (Wang et al., 2023), and many attempts raise the share of coding tasks solved by large factors (Brown et al., 2024). Spending inference computation deliberately can beat switching to a much larger model (Snell et al., 2024), thinking budgets can be controlled directly (Muennighoff et al., 2025), and models trained to reason at length have made extended deliberation a standard product feature (DeepSeek-AI, 2025), although on some tasks longer thinking makes results worse (Gema et al., 2025). Commercial coding agents expose this as a simple effort setting, with little published evidence of what it delivers in real agentic coding. The High versus xHigh contrast measures exactly that, and the prompt ablation shows one limit of the lever, since a design prompt consumed part of the reliability that effort otherwise bought.

### Cost, configuration, and measurement threats

Evaluations of agents must also account for what they cost. Ranking on accuracy alone encourages needlessly expensive designs (Kapoor et al., 2024). Resource limits reveal waste that accuracy numbers hide, such as token use that snowballs on failing runs (Fan et al., 2025). Consumption on the same task can differ by a factor of thirty, and spending more does not mean doing better (Bai et al., 2026), while tool calls carry an overhead that can outweigh what the tool contributes (Zhang et al., 2026). Evidence on AI assistance in real development is mixed. Developers were 56 percent faster in one controlled experiment (Peng et al., 2023) but 19 percent slower in a randomized trial with experienced maintainers who believed they had been faster (Becker et al., 2025). Early usability work found the tool did not straightforwardly improve success (Vaithilingam, Zhang & Glassman, 2022), and repository analysis links AI assisted speed to growing code complexity (He et al., 2026). Real teams configure agents through instruction files, tool grants, and settings (Galster et al., 2026), the surface this study manipulates. To the author's knowledge, no published study isolates what a browser based visual testing tool adds, in function and in tokens, across matched and repeated runs.

Two measurement problems shaped the design. Model performance is sensitive to details of the prompt that carry no meaning: formatting alone can move accuracy by tens of points (Sclar et al., 2024), rephrasing can change which model looks best (Mizrahi et al., 2024), and personas do not reliably help (Zheng et al., 2024). The prompt ablation confronts this by comparing a long third-party prompt against a one paragraph paraphrase of the part that matters. Output is also not repeatable. Setting temperature to zero does not guarantee identical answers, and only about a fifth of code generation papers account for this (Ouyang et al., 2025). Settings meant to be deterministic still vary (Atil et al., 2024), single run judgments are unreliable (Song et al., 2025), and reproductions of published LLM based software engineering studies mostly fail (Angermeir et al., 2025). This is why the study repeats identical configurations, changes one factor at a time, and reports run to run spread as a result in its own right.

## Materials and Methods

### Study design and reporting

The study is a matched comparison with repeated runs, organized as a grid. The factors are the model family and generation, the reasoning effort setting (High or xHigh, where the model offers it), whether the visual automation and testing tool (Playwright) is available, and the design-oriented system prompt (absent, full, or a one paragraph abridged form). Conditions were chosen rather than assigned at random, so the statistics in this paper describe the strength of patterns in the data rather than the outcome of a randomized experiment. The terms used for factors, treatments, and validity threats follow standard guidance for experiments in software engineering (Wohlin et al., 2012). Reporting follows the community guidelines for empirical studies involving large language models (Baltes et al., 2025), supplemented by the ACM SIGSOFT empirical standards (Ralph et al., 2020).

### Task and harness

Every run received the same specification, written as an OpenSpec change, for a self-hosted real time retrospective board with a React and Vite front end and a Node.js back end. The required features were board creation and listing with SQLite persistence, configurable columns, guest sign in by display name, drag and drop of cards between columns, nested comments, real time synchronization across connected clients over WebSockets, single container Docker deployment, CSV export, and developer documentation. All 85 Claude Code runs were produced with Claude Code CLI version 2.1.132, which was held at that version for the whole run window, using the /opsx:apply command to ask the agent to implement the OpenSpec specification. A subset of runs used the Antigravity agent version 1.23.2 rather than Claude Code. Prompting was zero shot: each agent received the specification through a single command invocation, /opsx:apply realtime-retro-board, with no worked examples of a solution and no pre or post processing of its output. The specification itself was produced before the first run, through the OpenSpec workflow's /opsx:explore and /opsx:propose commands, from the author's concept and requirements, and then frozen; every run folder holds a byte identical copy of its documents (hashes in Article S1). It was drafted with the same coding agent from the author's concept and hard requirements, and the author reviewed and edited the documents before they were frozen. OpenSpec version 1.3.1 was used for this study.

Models are identified by the product-facing versions exposed by the agent harnesses. Two harnesses (Claude Code CLI and Antigravity) were used, each with more than one model family: Claude Code CLI ran Claude Opus 4.7, Claude Opus 4.6, Claude Sonnet 4.6, Qwen 3.6 and Qwen Coder Next (85 runs), and the Antigravity agent ran Claude Opus 4.6, Claude Sonnet 4.6, Gemini 3.1 Pro at two effort settings and Gemini 3.1 Flash (5 runs). Claude Opus 4.6 and Claude Sonnet 4.6 were each run under both harnesses. The Antigravity runs are design prompted by construction, since the full design prompt is that agent's own injected system prompt. That harness does not expose the High and xHigh effort setting used in the Claude Code contrasts; its two Gemini 3.1 Pro runs used the agent's own high and low thinking levels, and its Claude runs record no effort setting. Runs were carried out from early May to early July 2026. Because hosted models are updated without a change to the product version, this limits reproducibility, as discussed under Limitations.

Each agent generated, built, and exercised its implementation within its own command execution environment during the run. On the Windows host used here that environment can be a Unix like shell or a Windows Subsystem for Linux instance rather than the native Windows shell, and it was not separately recorded. Rubric verification, including bringing up the local development server, building and running the Docker image, and confirming persistence across a restart, was carried out by the author on the native Windows host. Where these environments differed, an application that ran during the agent's session could still fail at verification through environment dependent faults, such as platform specific native modules or shell specific start scripts, and these cases were scored as failures of the relevant build or environment criterion.

### Models, conditions, and the design prompts

The dataset contains 90 graded runs, 85 through Claude Code CLI and 5 through the Antigravity agent. The Claude Code CLI runs cover Claude Opus 4.7 (48 runs), Claude Opus 4.6 (26 runs), Claude Sonnet 4.6 (9 runs), and two local Qwen models (Qwen 3.6 and Qwen Coder Next, one run each). The two Qwen models, Qwen 3.6 (27 billion parameters) and Qwen Coder Next (an 80 billion parameter mixture of experts model), ran locally in LM Studio on a single NVIDIA RTX PRO 6000 GPU (96 GB) under Windows, each as a Q4_K_M quantized build with a 262,144 token context window; Claude Code CLI was pointed at LM Studio's local server port in place of the hosted API, so the harness, the specification and the scoring were the same as for the hosted models. The Antigravity runs cover Claude Opus 4.6 and Claude Sonnet 4.6, one run each, and three Gemini 3.1 configurations, Pro at two effort settings and Flash. Within the Claude Code CLI runs, three factors were varied. The visual automation and testing tool, the Playwright MCP server, which exposes browser navigation, clicking, typing and screenshot tools to the agent, was either available or not. A design-oriented system prompt was either absent, applied in full, or applied in a one paragraph abridged form described below. Reasoning effort was set to High or, where the model supported it, to xHigh; one further Sonnet 4.6 run with the full design prompt used that model's Max setting and is counted among the full prompt runs but, having no matched partner, enters none of the effort contrasts. The Opus 4.7 family covers a full two by four grid, namely two effort levels by four conditions (base, plus Playwright, plus the full design prompt, plus the abridged design prompt), with six repeated runs in each of the eight cells. The base, Playwright, and full design prompt cells were run first, as a two by three effort sweep. The abridged prompt cells were added afterward as an ablation, so the pooled effort contrasts reported below are computed on the six sweep cells, and the abridged cells are reported separately. For every tool enabled run, the rubric records whether the agent actually invoked the testing functionality during its session. All 23 tool-enabled Claude Code runs did, so the tool contrasts below compare used tools, not merely available ones.

The full design-oriented prompt is the system prompt that the Antigravity agent, from Google, injects into its own sessions. It was recovered from that agent and, for the Claude Code runs, supplied unchanged as project level instructions in the project's CLAUDE.md file. It is a general-purpose coding agent prompt. The portion relevant here is a short design directive that tells the agent to treat visual quality as a first-class priority and to produce a premium, modern interface rather than a plain or minimal one, recommending modern design techniques in general terms, such as considered color and typography, gradients, motion, and dark themes. In what follows, the design directive means this instruction in either of its two forms, the design portion of the full prompt or the abridged paraphrase described below, and the design prompt means whichever file carried it into a run. The prompt contains no specification of the retrospective board itself, that is, no features, layout, content, or application specific visual details, and no functional requirements, all of which came from the shared OpenSpec change. As a third-party material it is not reproduced here, although its effect and general character are described.

To test whether the full prompt's length and emphatic tone are necessary, a one paragraph abridged directive was written for this study. It paraphrases the content of the design portion alone, dropping the rest of the agent prompt and its emphasis. Because it is a paraphrase rather than an excerpt, the abridged directive differs from the original in three ways at once: it uses different words, it is much shorter, and it drops the emphatic tone. The comparison therefore tests whether the shorter, plainer version produces the same effect; it cannot say which of the three differences is responsible, since none of them was varied on its own. Eighteen additional runs, six on Opus 4.6 at High effort and six each on Opus 4.7 at High and at xHigh effort, all with no testing tool, received it as the project's CLAUDE.md, under the same harness and Claude Code CLI version as the main batches. Unlike the third-party prompt, the abridged directive can be reproduced in full. It is reflowed here as one paragraph; the archived file, abridged_design_prompt.md, keeps the heading and bullet layout in which the runs received it:

> *Treat visual quality as a first-class priority for any user interface you build. The result should feel premium, modern, and polished at first glance, never plain or minimal. Use a curated, harmonious color palette instead of generic default colors; sleek dark themes are welcome. Use modern typography (for example a Google font such as Inter or Outfit) instead of browser defaults. Use smooth* *gradients and subtle micro-animations, and add hover effects so the interface feels responsive and alive. Do not ship a design that looks like a simple minimum viable product.*

### Functional verification and scoring

Each run was verified against the same 14 functional criteria. Scoring was based on running the software rather than on opinion: for each criterion the author performed the operation it describes on the delivered application, bringing up the local development server, building and running the Docker image, restarting the server to confirm persistence, moving cards between columns, opening comment threads, exporting CSV, and so on, and recorded what happened. The local development environment criterion asks whether the delivered project can be installed and started on the verification host by following its own instructions, typically npm install followed by npm run dev, without editing any file. A first try failure on this criterion therefore means that dependency installation failed or the development server did not start, so the application could not be exercised at all until the agent had been told the error. Most criteria therefore come down to something observable, a build succeeds or fails, data survives a restart or does not, a card moves or does not, rather than to a judgment call. Where a criterion does involve judgment, such as whether the developer documentation is adequate, it was scored against the items the specification requires. Each criterion received a 3 if the feature worked on the first try, a 2 if it failed at first and was fixed after a single corrective prompt, or a 1 if the fault was still present after that one corrective prompt. The corrective prompt described only the observed error or failing behavior and offered no guidance on how to fix it, so a score of 2 reflects the model's own ability to diagnose and repair the fault from the symptom alone. Corrective prompts addressed one issue at a time, so a single run could involve several of them when fixing one fault surfaced another, but each criterion received at most one, so a rating of 1 marks a fault the agent could not repair from a single description of its symptom. A run total is the sum of its 14 criterion scores, so the maximum is 42. The instrument was finalized at 14 criteria at dataset version 2.0.0, when a fifteenth criterion, data persistence across a container restart, was dropped because it measured how the container was launched rather than the delivered code; the retained criteria and the three point scale were unchanged, and the 22 runs graded before that release were re-scored under the final rubric, so every total reported here is on the same 42 point scale. Because a criterion rated 2 corresponds to exactly one corrective prompt, the count of criteria rated 2 in a run is used here as a per run measure of repair burden, that is, a lower bound on the number of human interventions the run required.

All verification was performed by the author, who knew each run's configuration while verifying it. What this means for the results is discussed under limitations. The per run rubric files preserve the recorded outcome of every criterion, and the archived artifacts let anyone re-check any criterion, so the scoring can be audited rather than taken on trust.

### Measuring the design treatment

The design treatment was measured directly from each run's shipped source, without a judge. A script archived with the dataset (count_design_markers.py) scans every run's CSS, TSX, JSX, TS, JS and HTML files, excluding dependencies and build output, and counts two families of features. The first is what the design directive names: CSS gradients, keyframe animations, Google Fonts imports, and named display faces such as Inter or Outfit. The second is styling effort the directive does not name: lines of CSS, CSS custom properties (design tokens), box-shadow, border-radius, transform and letter-spacing rules, the number of distinct font sizes and weights, and the number of distinct dark colours. A run is classed as design-treated if it contains at least one gradient and either a Google Fonts import or a keyframe animation. This rule is only a convenient label: as Table 6 shows, every prompted run contains all four named features and no unprompted run contains any of them, so any one feature alone would classify the runs in exactly the same way. These measures are the primary evidence for every claim about the design prompt in this paper. They capture how much, and what kind of, design work the agent shipped, and nothing about whether the result looks good; that judgment is left to the reader, from Figure 5 and the archived screenshots. The per run measures are listed in Table S3.

A holistic visual rating is reported alongside as a secondary observation. It was produced on a 1 to 5 scale from each run's dashboard screenshots by a large language model, Claude Opus 4.8 accessed through the Claude desktop application, asked to judge polish and finish, color and typography coherence, layout and visual hierarchy, and overall professional feel; the author then went through every rating. The model generated the descriptive scale anchors itself rather than being given them, and it saw the run folder names, which encode the condition, so the rating was not blinded. Rating interfaces from screenshots with a model has precedent in front end evaluation (Si et al., 2025; Zhang et al., 2025) and known weaknesses: model raters agree only partly with human judgments of interfaces (Luera et al., 2025), carry the general biases of model judges (Zheng et al., 2023), and can favor work produced by their own model family (Panickssery, Bowman & Feng, 2024). For these reasons no conclusion rests on the rating and no inferential claim is built on it. Its role is to be compared with the source-based measures: the results report the agreement between the two, including within the prompted runs, and all 90 screenshots are archived with the dataset and the anchors as applied are given in Article S3, so that any reader can form an independent view. Every run, with its harness and configuration, raw functional score, the 14 criterion ratings, session cost, base-10 logarithm of cost, and visual rating, is listed in Table S1.

## Analysis

Per run session cost is right skewed, because a single difficult debugging path can double the cost of a run. Cost is therefore summarized with the median, ranges are reported, and the median is used when comparing conditions. All cost figures are the session cost reported by Claude Code CLI's /cost command, which estimates a dollar amount locally from token counts and per-token model pricing, so a local model's figure reflects token volume at hosted-model rates rather than its actual running cost. Alongside cost, each run's rubric file preserves the session summary reported by the harness at the end of the run: cumulative model time, lines of code added and removed, and token usage for the primary model split into input, output, and cache read tokens. Where a rubric preserves both an initial and a final summary, the final one is used, because it is cumulative over the whole session. These records support the token level cost decomposition reported in the results. Functional scores are reported as means, medians, and ranges. Figure 2 shows every run as a point, with the group mean marked for scores, which pile up at the ceiling, and the group median for cost, which is right skewed.

Where a contrast between two groups of runs is summarized inferentially, a two-sided Fisher exact test on run counts is reported, and for the key comparisons of proportions the difference between the two rates is reported with a 95 percent confidence interval (Newcombe's method), which gives the range of differences the data are consistent with. Every inferential comparison in this paper is listed with the counts behind it in Table S2, and the derivation of each from the archived rubric files is supplied as Article S1, generated by the derivation script archived with the dataset. Because conditions were chosen rather than randomized, these p values and intervals are descriptive summaries of the strength of association rather than tests of randomized hypotheses. The Spearman rank correlations and two-sided Mann-Whitney U tests that compare the holistic visual rating with the source-based design measures are descriptive in the same sense: they summarize how closely the two agree, and they do not test a hypothesis. Effects are otherwise reported descriptively, and no causal identification is claimed beyond the matched contrasts. Small cells also limit what the study can detect: with six to seven runs per cell, only near categorical differences reach conventional significance, and differences in run totals smaller than about two rubric points would often be missed. A null result in this paper therefore means that no effect large enough to matter here was found, not that no effect of any size exists. The dataset, the per run rubric files, and the scoring are openly available, as described under data availability.

## Results

### Capability tier dominates the outcome

The single largest source of variation in functional score was the model family, not any tool, prompt, or effort setting. Among the three frontier families the scores clustered near the ceiling, with family means of about 41 of 42 for Opus 4.7, Opus 4.6, and Sonnet 4.6. These runs spanned a wide cost range, from about $1 to $8 each, yet spending more within this band did not buy a higher score. The score differences that tools, prompts, or effort produced inside a tier were on the order of one to two points, although, as the next section shows, these compressed totals conceal much larger differences in first try reliability.

Two low-cost local models, Qwen 3.6 and Qwen Coder Next, sat in a different regime on both axes, scoring 37 and 24 respectively, roughly ten points below the frontier band. The session cost reported by Claude Code CLI's /cost command was $41 and $179, against a frontier median of about $3.60 and a frontier maximum of $7.63 (Article S1), so on the order of twelve to fifty times the frontier median and more than five times the most expensive frontier run. These two dollar figures are estimates rather than prices actually paid. The /cost command estimates a dollar amount from token counts and per-token model pricing. For the hosted frontier runs that approximates the real price, but Qwen ran locally, where inference is essentially free and no such price applies. Its value is therefore best read as the token volume the run consumed, priced at hosted-model rates rather than as a market cost, with the exact rate applied to the local model undocumented, and the large figure reflects the heavy iteration the weaker model required. The gap to this model, roughly ten points in score, is far larger than any difference that tools, prompts, or effort produced within a tier. Because the Gemini and Qwen families contribute only three and two runs, their rows in Table 1 are exploratory context rather than family comparisons. Table 1 summarizes the families and Figure 1 shows score against cost on a logarithmic cost axis.

**Table 1.** Dataset summary by harness and model family. The Antigravity harness does not report session cost, so cost is unavailable for those five runs, and every cost range shown is computed over the runs in its own row. The Gemini and Qwen rows are exploratory (three and two runs) and are not comparable family means. The Qwen cost range is a token volume estimate at hosted model rates for locally run inference, not a price.

| **Model family** | **Runs** | **Mean score** | **Score range** | **Cost range (USD)** |
|---|---|---|---|---|
| **Claude Code harness** | | | | |
| Claude Opus 4.7 | 48 | 41.3 | 39 to 42 | $2.22 to $7.63 |
| Claude Opus 4.6 | 26 | 40.7 | 38 to 42 | $2.04 to $6.62 |
| Claude Sonnet 4.6 | 9 | 40.9 | 39 to 42 | $1.08 to $4.90 |
| Qwen (3.6, Coder Next) | 2 | 30.5 | 24 to 37 | $41 to $179 |
| **Antigravity harness** | | | | |
| Claude Opus 4.6 | 1 | 40.0 | 40 | not recorded |
| Claude Sonnet 4.6 | 1 | 42.0 | 42 | not recorded |
| Gemini 3.1 (Pro high, Pro low, Flash) | 3 | 39.3 | 38 to 40 | not recorded |

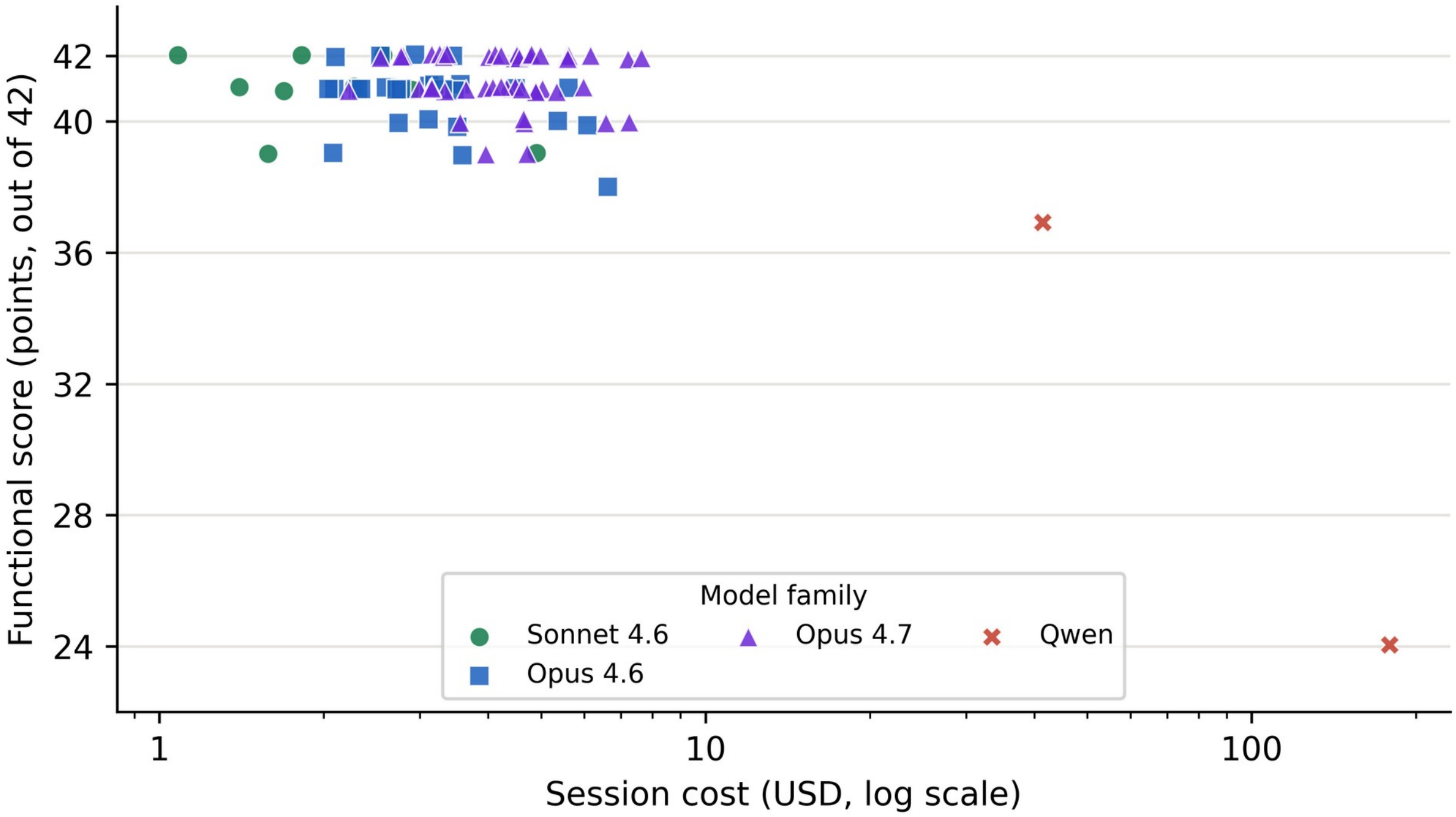

**Figure 1.** Functional score against session cost (logarithmic cost axis), with model families distinguished by marker shape and color. Frontier models cluster near the 42 point ceiling across a wide cost range. The local model (Qwen), whose plotted value is a token-usage-based estimate from Claude Code's /cost command rather than a market price, sits far below and far to the right. The two Qwen runs are context, not a family comparison. Scores are jittered vertically for visibility, and the five Antigravity runs, which have no recorded cost, are not shown (n = 85).

### Run totals conceal how first-try reliability differs within the frontier

Within the frontier, models that look interchangeable on post repair totals differ sharply in what they deliver on the first attempt. Because a failed criterion that is repaired after one corrective prompt still earns 2 of its 3 points, the 42 point total is dominated by what a model can eventually fix rather than by what it gets right the first time. Re-reading the same runs at the criterion level shows differences within the frontier that the totals compress to a point or less (Table 2). At High effort, pooling each family's High effort Claude Code runs, only 4 of 26 Opus 4.6 runs were perfect on the first try, against 6 of 24 for Opus 4.7 and 3 of 8 for Sonnet 4.6. These pooled family counts include the abridged prompt cells, which Table 2 also breaks out as separate rows. The sharpest movement was on the single dominant defect: the Docker deployment criterion passed first try in only 5 of 26 Opus 4.6 High runs but in 18 of 24 Opus 4.7 High runs (Fisher exact $p = 0.00016$; risk difference 56 percentage points, 95 percent confidence interval 28 to 73; Table S2), a swing invisible in family means that differ by less than one point (40.7 versus 41.3).

The generation change did not remove environment failures so much as move them. While Docker first try passes rose sharply from Opus 4.6 to Opus 4.7 at High effort, first try failures on the local development environment criterion went the other way, from 1 of 26 runs to 10 of 24 (Table 2, pooling each family's High rows), largely package installation faults surfacing before the development server came up. Only raising reasoning effort cleared both at once in the sweep: at xHigh, 17 of 18 runs passed Docker first try and all 18 brought the local environment up first try (Table 2), although, as the ablation below shows, a design prompt can reintroduce these faults even at xHigh. The practical reading is that a newer generation reshuffles which environment faults occur, whereas added deliberation is what eliminates them, and that first try reliability and repair burden are more informative reporting axes than post repair totals for comparing models near a rubric ceiling.

**Table 2.** First-try reliability and repair burden by configuration. Claude Code runs only. The five Antigravity harness runs and a single Sonnet 4.6 run at Max effort are excluded. First-try perfect means all 14 criteria rated 3, with no corrective prompt. Corrective prompts per run counts criteria rated 2. Local environment first-try pass counts criterion 1 rated 3.

| Configuration | Runs | First-try perfect | Docker first-try pass | Local environment first-try pass | Corrective prompts per run (mean) |
|---|---|---|---|---|---|
| Opus 4.6, High | 20 | 1 of 20 (5%) | 2 of 20 | 19 of 20 | 1.25 |
| Opus 4.6, High, abridged prompt | 6 | 3 of 6 (50%) | 3 of 6 | 6 of 6 | 0.83 |
| Sonnet 4.6, High | 8 | 3 of 8 (38%) | 4 of 8 | 7 of 8 | 0.88 |
| Opus 4.7, High | 18 | 5 of 18 (28%) | 13 of 18 | 10 of 18 | 0.89 |
| Opus 4.7, High, abridged prompt | 6 | 1 of 6 (17%) | 5 of 6 | 4 of 6 | 1.17 |
| Opus 4.7, xHigh | 18 | 16 of 18 (89%) | 17 of 18 | 18 of 18 | 0.17 |
| Opus 4.7, xHigh, abridged prompt | 6 | 1 of 6 (17%) | 3 of 6 | 5 of 6 | 1.00 |
| Qwen, High | 2 | 0 of 2 (0%) | 0 of 2 | 1 of 2 | 4.50 |

### Where the failures were: a criterion level profile

Across all 90 runs, first try failures were heavily concentrated in the environment criteria (Table 3). The Docker deployment criterion failed on the first attempt in 40 of 90 runs, 44 percent, and the local development environment criterion in 15. The rubric notes attribute those fifteen to a failed dependency installation (npm install) in nine runs, a development server that would not start (npm run dev) in three, and in one run a native SQLite module that had to be replaced before the project would build; the remaining two notes record only the failure. The two environment criteria account for 55 of the 100 first try criterion failures in the dataset, 55 percent. The recurring culprits were a native SQLite module (better-sqlite3) that does not compile in a minimal Alpine based container, which runs that chose a WebAssembly database avoided, and a breaking change in a web framework (Express 5, whose routing library no longer accepts the bare wildcard route app.get('*') that the agents used to serve the front end). The largest interface side contributor for failures was moving cards between columns, the drag and drop criterion, at 17 runs, 19 percent, a concentration examined further in the design prompt section. Every other interface criterion failed in 6 percent of runs or fewer. Notably, no run left Docker unresolved after corrective prompting, whereas 7 of the 11 criteria left unresolved anywhere in the dataset belong to the two Qwen runs, so criteria rated 1 are almost entirely a weak model phenomenon rather than a task phenomenon.

**Table 3.** First-try failure profile across all 90 runs. Rated 2 means failed first try and fixed after one corrective prompt. Rated 1 means never fully resolved. The share column is the fraction of the 90 runs that missed the criterion on the first attempt.

| Criterion | Rated 2 | Rated 1 | First-try failure share |
|---|---|---|---|
| Local development environment | 15 | 0 | 17% |
| Docker deployment | 40 | 0 | 44% |
| Home page | 2 | 0 | 2% |
| Board creation | 5 | 0 | 6% |
| User identification | 1 | 1 | 2% |
| Card interaction | 3 | 0 | 3% |
| Moving cards between columns | 14 | 3 | 19% |
| Commenting | 3 | 2 | 6% |
| Realtime update: new card | 2 | 1 | 3% |
| Realtime update: card move | 1 | 1 | 2% |
| Realtime update: new comment | 0 | 1 | 1% |
| Data persistence across restart | 0 | 1 | 1% |
| Documentation | 2 | 0 | 2% |
| CSV export | 1 | 1 | 2% |

## The testing tool added cost, not quality or reliability

Turning the Playwright tool on, with model and effort held fixed, did not change the functional score, but it did raise cost. On Opus 4.7 the median session cost rose by 42 percent at High effort and by 68 percent at xHigh effort, while the functional score was unchanged (Table 5; Figure 2). The same pattern appeared in the earlier Opus 4.6 contrast, where seven runs with the tool and seven without it produced the same mean functional score of 40.6 while the tool raised the median cost by 27 percent (Article S1). Notably, several tool enabled runs still failed on the Docker build, a fault the tool cannot reach, because exercising the running interface surfaces user interface problems but not the build, runtime configuration, or data persistence faults beneath it, and when the build itself fails there is no interface to test. Those build and environment faults were where most failures actually occurred (Table 3), which explains why a tool aimed at the rendered page did not improve reliability.

In the pooled distributions plotted in Figure 2, twelve runs per group, the group means are 41.50 without the tool and 41.75 with it, and the medians are 42 in both. That difference is produced almost entirely by one base run scoring 39, and excluding it leaves a difference in means of 0.02 points. The gap is an artifact of a single low run rather than a systematic lift, and in either case it falls far below the roughly two point difference this design can resolve.

Two further checks close off the easy objections to this null result. First, the tool was not simply ignored: every tool enabled run's rubric records the testing functionality as actually invoked during the session. Second, the null is not explained by a lack of interface visible faults to catch. In the matched Opus 4.6 contrast, where interface visible faults did occur and the tool therefore had defects available to catch, first try failures on interface visible criteria occurred in 1 of 7 runs without the tool and 3 of 7 with it, and tool enabled runs elsewhere in the dataset still failed home page rendering, commenting, and drag and drop criteria first try. Within this dataset, passive screenshot verification did not reduce even the class of defect it is designed to see. A third consideration bounds the claim rather than the result: because the cells are small (see Analysis), the finding is that the tool bought no improvement of practical size while raising median cost by 42 to 68 percent on Opus 4.7 and 27 percent on Opus 4.6, a cost without payoff pattern that matches the tool overheads reported for LLM agents more broadly (Kapoor et al., 2024; Fan et al., 2025).

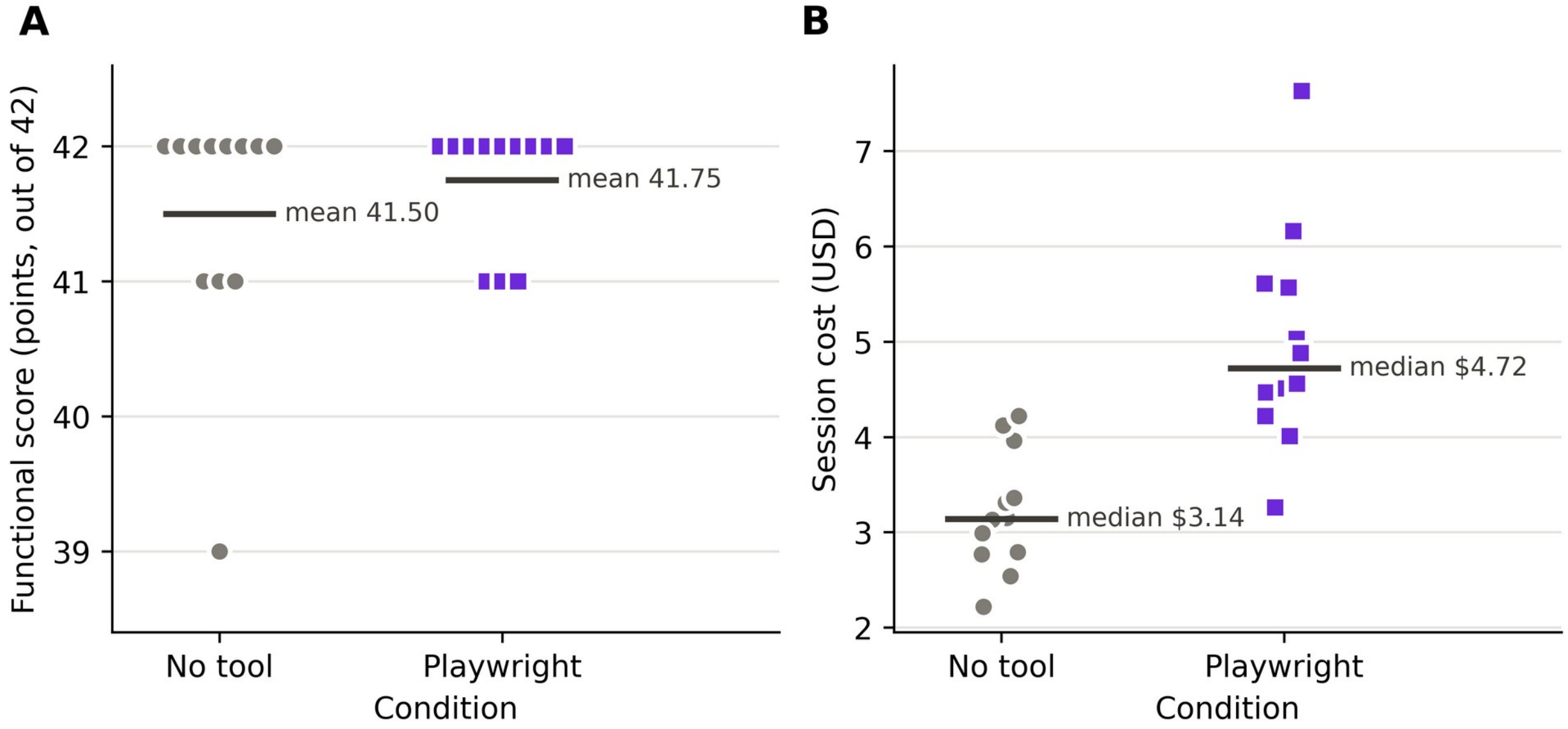

**Figure 2.** Effect of adding the testing tool on Opus 4.7, pooling High and xHigh effort, with twelve runs in each group. Each dot is one run, and identical scores are fanned out horizontally so repeated values remain visible. (A) Functional score. Three runs scored 41 in each group and the rest scored 42, eight without the tool and nine with it. The only other run is a single no tool run at 39, which accounts for almost all of the gap between the group means of 41.50 and 41.75 marked by the horizontal lines. Excluding it leaves a difference of 0.02 points. (B) Session cost, with the group median marked, rising from $3.14 without the tool to $4.72 with it. The summary line is the mean in panel A, where scores pile up against the 42 point ceiling and the median carries no information, and the median in panel B, where cost is right skewed.

The session token records show what each premium actually buys (Table 4). With the tool, median output tokens are flat or slightly lower at High, 42k versus 46k, and only modestly higher at xHigh, 51k versus 45k, while median cache read tokens rise two to three-fold, from 2.3M to 5.3M at High and from 2.3M to 7.0M at xHigh. The tool premium is therefore context re-reading, the model repeatedly re-ingesting a session inflated by tool roundtrips and screenshots, not additional engineering output. Recent work calls this kind of overhead a tool use tax, a cost of using the tool itself rather than of the work the tool enables (Zhang et al., 2026). The design prompts have the opposite signature: median output tokens rise over base, by about a quarter at High, 55k to 60k versus 46k, and by more than half at xHigh, 72k versus 45k, and the median lines of code added rise from about 2,200 to between 3,000 and 3,900, because the agent writes substantially more interface code. The full and abridged prompts are indistinguishable on this axis, which confirms that the premium follows the interface work the directive elicits rather than the prompt text itself. The additions cost more for entirely different reasons, and the median functional score is essentially unchanged in every cell.

**Table 4.** Token, time, and code decomposition of the Opus 4.7 grid, medians per cell of six runs. Output and cache read tokens are the primary model's counts from the end of session report. Model time is cumulative model API time. Lines added is the harness's count of code lines added over the session. The xHigh base token medians are over the five runs in that cell whose rubric preserves the per model token line.

| Configuration | Score (median) | Output tokens (median) | Cache-read tokens (median) | Model time, min (median) | Lines added (median) | Cost (median, USD) |
|---|---|---|---|---|---|---|
| High, base | 41 | 46k | 2.3M | 9.1 | 2,200 | $3.06 |
| High, plus Playwright | 41.5 | 42k | 5.3M | 9.6 | 1,804 | $4.34 |
| High, plus full design prompt | 41 | 55k | 3.8M | 11.1 | 3,011 | $4.28 |
| High, plus abridged prompt | 41 | 60k | 3.9M | 11.3 | 3,268 | $4.14 |
| xHigh, base | 42 | 45k | 2.3M | 9.2 | 2,150 | $3.33 |
| xHigh, plus Playwright | 42 | 51k | 7.0M | 11.9 | 2,458 | $5.59 |
| xHigh, plus full design prompt | 42 | 72k | 5.2M | 14.4 | 3,878 | $5.17 |
| xHigh, plus abridged prompt | 41 | 72k | 4.2M | 14.3 | 3,620 | $4.73 |

## Reasoning effort buys first-try reliability

The reliability that the tool did not provide was available from reasoning effort. Pooling the Opus 4.7 sweep conditions, raising effort from High to xHigh raised the share of runs that were perfect on the first try, meaning a score of 42 with no corrective prompts, from 28 percent (5 of 18) to 89 percent (16 of 18), a contrast that is large by any standard (Fisher exact $p = 0.00049$; risk difference 61 percentage points, 95 percent confidence interval 29 to 78; Table S2). Figure 3 breaks the first try rate out by condition, and the gain appears in every sweep cell. The pooled mean score rose from 41.0 to 41.8 (Article S1). In repair burden terms, the 18 High runs required 16 corrective prompts between them and the 18 xHigh runs required 3, a mean of 0.89 against 0.17 per run (Table 2) and about a five-fold reduction in human interventions. The High-level misses were concentrated in first run environment failures, namely npm install, Docker build and run, and one case of data loss on restart, which are exactly the kind of defect that more deliberation catches before the work is handed back. The cost of this gain was modest. Holding the condition fixed, the median cost premium for xHigh was about 9 percent for base runs, 29 percent with the tool, and 21 percent with the design prompt (Table 5). In the base condition the xHigh runs emitted essentially the same output tokens as the High runs (Table 4), so the extra deliberation is not simply additional generation. Table 5 reports the full Opus 4.7 grid. The pooled contrasts above use the six sweep cells, and the abridged prompt cells are taken up in the ablation below.

**Table 5.** Opus 4.7 effort grid. Eight configurations, six repeated runs each. The two abridged prompt cells belong to the ablation, are excluded from the pooled effort contrasts in the text, and reappear with the Opus 4.6 ablation cells in Table 7. The visual rating column gives the number of runs rated 3, 4 and 5; no run in these cells was rated 1 or 2.

| Configuration | Mean score | First-try perfect | Median cost (USD) | Visual rating (runs at 3 / 4 / 5) |
|---|---|---|---|---|
| High, base | 41.0 | 2 of 6 | $3.06 | 6 / 0 / 0 |
| High, plus Playwright | 41.5 | 3 of 6 | $4.34 | 6 / 0 / 0 |
| High, plus full design prompt | 40.5 | 0 of 6 | $4.28 | 0 / 1 / 5 |
| High, plus abridged prompt | 40.8 | 1 of 6 | $4.14 | 0 / 0 / 6 |
| xHigh, base | 42.0 | 6 of 6 | $3.33 | 6 / 0 / 0 |
| xHigh, plus Playwright | 42.0 | 6 of 6 | $5.59 | 6 / 0 / 0 |
| xHigh, plus full design prompt | 41.5 | 4 of 6 | $5.17 | 0 / 1 / 5 |
| xHigh, plus abridged prompt | 41.0 | 1 of 6 | $4.73 | 0 / 0 / 6 |

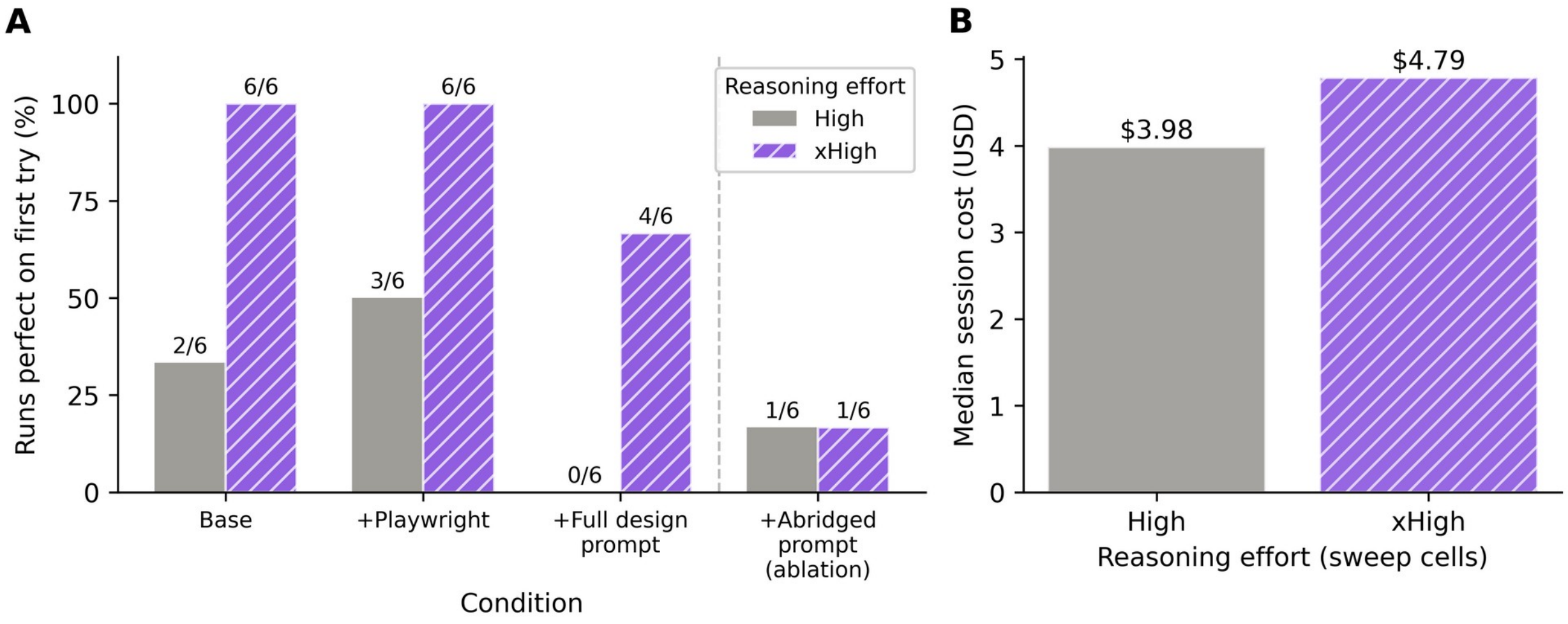

**Figure 3.** Reasoning effort on Opus 4.7, High versus xHigh. (A) The share of runs perfect on the first try, by condition. Raising effort lifts the base and Playwright cells to the ceiling and partially recovers the full design prompt cell, but leaves the abridged prompt cell unchanged. The dashed line separates the sweep conditions from the abridged prompt ablation, which is excluded from the pooled statistics in the text. (B) Median session cost across the pooled sweep runs rises modestly.

### The design prompt lifts visual quality, not function

The design-oriented prompt changed what the agents built, and the change was categorical. Among the 85 Claude Code runs, none of the 45 without a design prompt shipped a CSS gradient, a keyframe animation, a Google Fonts import or a named display face; all 22 with the full prompt and all 18 with the abridged paraphrase shipped all four (Table 6). The difference was not confined to the techniques the directive names. Prompted runs defined about three times as many design tokens (median 33 and 30 custom properties against 10, with no overlap between the groups), wrote about three times as much CSS (967 and 956 lines against 330), and used shadows, transforms and letter-spacing that the unprompted runs almost never used (transform rules, median 14 and 18 against 0; Table 6). No run in any condition depended on a UI or styling framework, so the unprompted runs did not achieve the same result by another route; they did not attempt it. Functional scores were indistinguishable from base within run-to-run variation and, if anything, drifted slightly lower, consistent with the prompt pushing the agent toward a more ambitious interface that occasionally costs a point. Nothing else moved the treatment: base runs carried none of the markers at either effort level, with or without the testing tool, and across Opus 4.6, Opus 4.7, and Sonnet 4.6 alike, so it is the design prompt rather than raw capability or compute that changed what was built. Figure 5 shows the underlying screenshots so that readers can judge the result directly.

The holistic visual rating agreed with the source-based classification in all 85 Claude Code runs and in 89 of 90 overall: every run classed as design-treated was rated 4 or 5 and every other Claude Code run was rated 3 (Figure 4), the single exception being the Antigravity Opus 4.6 run, which carries gradient and font markers but was rated 3. Within the 40 prompted runs, where the folder name carries no information the rater could use, runs rated 5 (n = 24) contained more of nearly every measure than runs rated 4 (n = 16): gradients 19 against 6, CSS lines 1,052 against 844, design tokens 34 against 28, shadows 16 against 9, transforms 19 against 12 and letter-spacing rules 11 against 3 (two-sided Mann–Whitney p of 0.004 or less for each; Article S1, from the per run values in Table S3), while distinct font sizes did not differ. Across the 85 runs the rating's Spearman correlation with a composite rank of eight source measures (gradient and keyframe counts, CSS lines, design tokens, shadows, transforms, distinct font sizes and dark colours) was 0.88, and 0.59 within the prompted runs alone (Article S1). The rater saw screenshots, not source, so its gradations tracked properties of files it never read. That does not make the rating blinded, but it bounds what the non-blinding could have done: no conclusion here rests on the rating, and where it is reported it agrees with an independent, judge-free measure. The remaining five runs, produced by the Antigravity agent through a different harness and model set, are listed in Table S1 and Table S3 and are kept out of the contrast because they vary the harness as well as the prompt. Three of the five carry a gradient and two meet the design-treated rule, and four of the five were rated 3 or below, so the association between the prompt and a designed interface did not survive the change of harness.

**Table 6.** Design treatment in the shipped source, by condition. Counted from each run's own CSS, TSX, JSX, TS, JS and HTML files, excluding dependencies and build output, by the archived script count_design_markers.py. The first block lists features the design directive names, as the number of runs containing at least one; a run is classed as design-treated if it contains a gradient and either a Google Fonts import or a keyframe animation. The second block lists technique-agnostic measures of styling effort that the directive does not name, as the median [minimum to maximum] per run. The last rows give the holistic visual rating for comparison. Font sizes and weights count distinct literal values, and dark colours are distinct hex colours with relative luminance below 0.15. No run in any condition depends on a UI or styling framework.

| Measure | No design prompt | Full design prompt | Abridged prompt | Antigravity harness |
|---|---|---|---|---|
| Runs | 45 | 22 | 18 | 5 |
| **Directive-named features (runs with at least one)** | | | | |
| CSS gradient | 0 of 45 | 22 of 22 | 18 of 18 | 3 of 5 |
| Keyframe animation | 0 of 45 | 22 of 22 | 18 of 18 | 2 of 5 |
| Google Fonts import | 0 of 45 | 22 of 22 | 18 of 18 | 2 of 5 |
| Named display font | 0 of 45 | 22 of 22 | 18 of 18 | 4 of 5 |
| Design-treated (rule above) | 0 of 45 | 22 of 22 | 18 of 18 | 2 of 5 |
| UI or styling framework dependency | 0 of 45 | 0 of 22 | 0 of 18 | 1 of 5 |
| **Styling effort not named by the directive (median [min to max] per run)** | | | | |
| CSS lines | 330 [0 to 805] | 967 [653 to 1591] | 956 [105 to 1376] | 297 [251 to 1167] |
| CSS custom properties (design tokens) | 10 [0 to 17] | 33 [22 to 57] | 30 [23 to 50] | 13 [7 to 37] |
| box-shadow rules | 3 [0 to 8] | 12 [6 to 22] | 15 [6 to 22] | 3 [2 to 12] |

| Measure | No design prompt | Full design prompt | Abridged prompt | Antigravity harness |
|---|---|---|---|---|
| border-radius rules | 12 [4 to 25] | 27 [17 to 40] | 31 [23 to 38] | 6 [4 to 32] |
| transform rules | 0 [0 to 3] | 14 [9 to 26] | 18 [3 to 26] | 3 [2 to 15] |
| letter-spacing rules | 0 [0 to 3] | 6 [2 to 18] | 10 [2 to 15] | 3 [0 to 5] |
| Distinct font weights | 2 [0 to 3] | 4 [3 to 5] | 4 [3 to 4] | 1 [1 to 4] |
| Distinct font sizes | 6 [0 to 15] | 14 [11 to 22] | 12 [0 to 23] | 8 [1 to 16] |
| Distinct dark colours | 1 [0 to 3] | 3 [0 to 6] | 4 [3 to 7] | 3 [0 to 4] |
| **Holistic visual rating (for comparison)** | | | | |
| Rated 4 or 5 | 0 of 45 | 22 of 22 | 18 of 18 | 1 of 5 |
| Rated 3 or below | 45 of 45 | 0 of 22 | 0 of 18 | 4 of 5 |

The criterion level data also locates the prompt’s occasional functional cost. At High effort the full design prompt roughly doubled the repair burden, 1.50 corrective prompts per run against 0.67 for base (Article S1), and its misses cluster on the most interaction heavy criterion: across all 90 runs, moving cards between columns failed first try in 11 of the 40 design prompted Claude Code runs, full or abridged, against 5 of the 45 runs without a design prompt, and the only two Opus 4.7 xHigh runs in the whole sweep that were not first try perfect were design prompt runs that both missed that drag and drop criterion, and one of them also failed the Docker build first try. Given the small counts this concentration is descriptive rather than definitive (Fisher exact $p = 0.093$; risk difference 16 percentage points, 95 percent confidence interval −0.4 to 33; Table S2), but it is consistent with the mechanism visible in the outputs: the prompt induces custom, animated interfaces whose richer interaction code is easier to get subtly wrong than the plain defaults. The abridged prompt runs supplied direct evidence of the mechanism: in one Opus 4.7 run the agent’s own diagnosis traced both a broken drag and a blurred comment panel to a backdrop filter blur it had added for visual depth, exactly the class of decorative effect the prompt encourages.

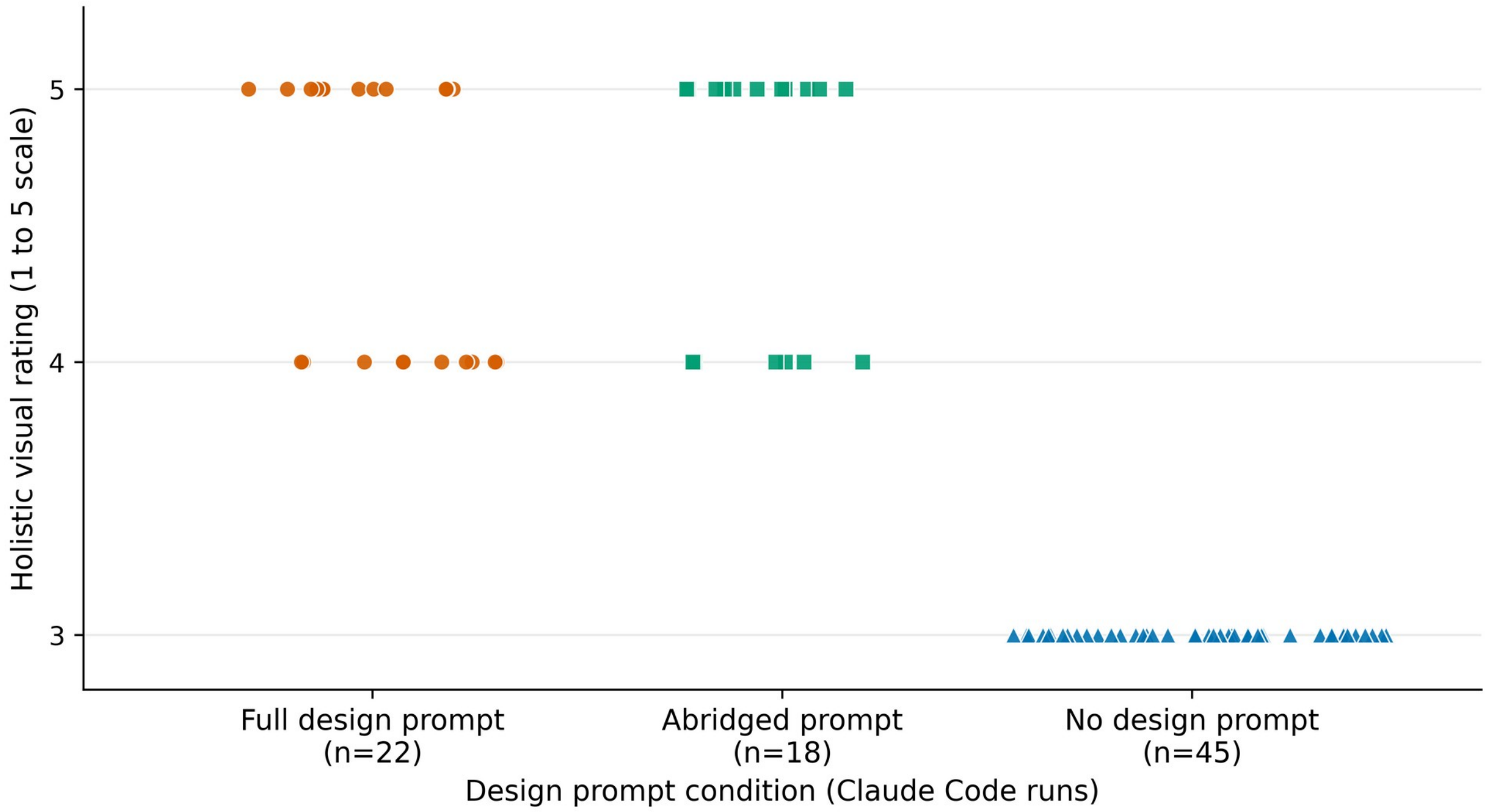

**Figure 4.** Holistic visual rating for the 85 Claude Code runs: full design prompt (n=22), abridged design prompt (n=18), and no design prompt (n=45). Each small dot is one run, jittered horizontally for visibility; no summary statistic is drawn, since the rating is ordinal and descriptive. All 45 unprompted runs sit at exactly 3 and all 40 prompted runs at 4 or 5, matching the source-based classification in Table 6 run for run, and the separation holds across reasoning effort, the testing tool, and model family. The vertical axis is restricted to the 3 to 5 range to show the clusters. The ratings were not blinded to condition.

Figure 5 makes the contrast concrete. Each row is one Opus configuration and each column is a condition: the base prompt, the base prompt plus the testing tool, the base prompt plus the full design prompt, and, from the ablation, the base prompt plus the abridged design prompt. The first two columns, which lack a design prompt, show clean but plain default interfaces across all three configurations. Adding the testing tool does not change the visual style. The two right hand columns, with the full and the abridged design prompt, show marketing style landing pages with custom product names, gradient headlines, and color-coded columns, regardless of model version or effort, and the two prompt forms are visually indistinguishable from each other. The jump from the left half to the right half is large and consistent, while moving down the rows, which raises capability and effort, leaves the visual style unchanged.

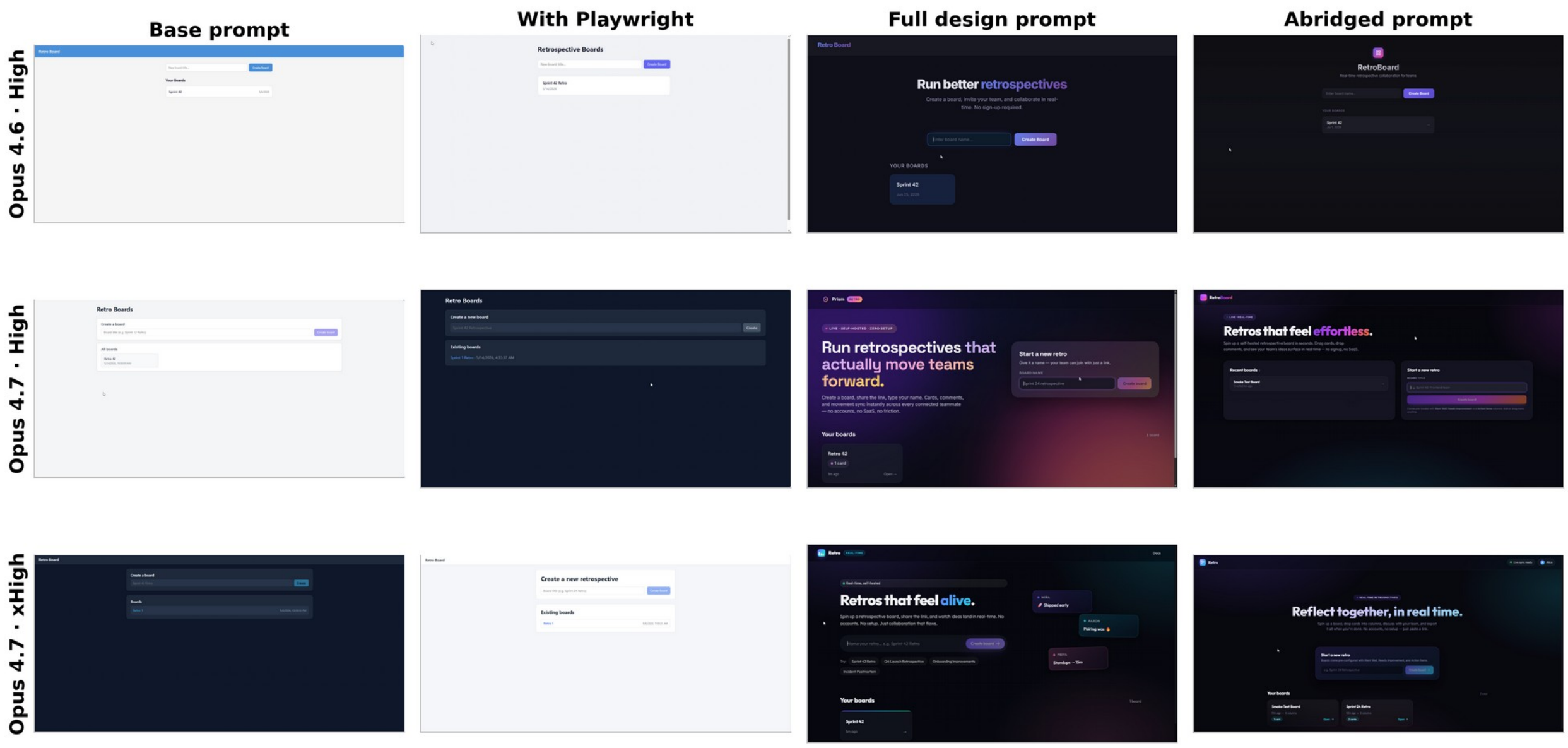

**Figure 5.** Dashboard screenshots for three Opus configurations (rows) under four conditions (columns): the base prompt, the base prompt with the testing tool, the base prompt with the full design prompt, and the base prompt with the abridged design prompt from the ablation. Neither the model version, the effort level, nor the testing tool changes the visual style. Either form of the design prompt does, and the two forms are indistinguishable. The runs shown are, for Opus 4.6 High, claude_opus_4.6_high, claude_opus_4.6_with_playwright_high, claude_opus_4.6_high_with_antigravity_prompt_run_5 and claude_opus_4.6_high_abridged_prompt; for Opus 4.7 High, claude_opus_4.7_high, claude_opus_4.7_with_playwright_high, claude_opus_4.7_high_with_antigravity_prompt_run_3 and claude_opus_4.7_high_abridged_prompt; and for Opus 4.7 xHigh, claude_opus_4.7_xhigh, claude_opus_4.7_with_playwright_xhigh, claude_opus_4.7_xhigh_with_antigravity_prompt_run_2 and claude_opus_4.7_xhigh_abridged_prompt. Every run in each of the twelve cells carries the same Table 6 classification as the run shown, so the choice of run affects the detail of a panel but not the contrast; the dashboard screenshots of all 90 runs are archived with the dataset.

## A one paragraph abridged prompt reproduces the lift

A one paragraph paraphrase delivers the same design treatment as the full third-party prompt, and the functional cost that design prompts carry persists even at higher effort. The comparison asked first whether the full prompt is necessary for the design effect, or whether the paraphrase suffices. It suffices (Table 7; Figure 5, two right hand columns). Every abridged run met all four source markers, with design token and CSS volumes indistinguishable from the full prompt cells (Table 6). On Opus 4.7 every abridged prompt run was rated 5 and the full prompt runs 4 or 5; on Opus 4.6 both groups were rated 4 or above. Every abridged run produced the same designed character the full prompt produces, dark themed, gradient accented interfaces on Opus 4.6 and complete marketing style landing pages with custom product names and gradient display headlines on Opus 4.7, while every base run in the dataset remained at 3. The functional pattern also carried over: abridged prompt means stayed within the frontier band (41.2 and 40.8), and the drag and drop concentration reappeared, with 3 of the 6 Opus 4.7 abridged runs missing the moving cards criterion first try. The cost premium was likewise similar to the full prompt's, a median of $4.14 versus $4.28 on Opus 4.7, against $3.06 for base (Table 7), indicating the premium pays for the additional interface code the directive elicits rather than for the prompt text itself. Since the abridged directive contains no emphasis, no threat of failure, and none of the surrounding agent instructions, the effect evidently depends on the content of the ask alone. Neither the third party prompt's length nor its insistent tone is necessary.

Extending the ablation to xHigh on Opus 4.7 asked a second question: does higher reasoning effort remove the functional cost that design prompts carry? It does not. The six abridged xHigh runs kept the full design treatment, every run rated 5, but only 1 of 6 was first try perfect against 6 of 6 for xHigh base (Table 7), with a mean score of 41.0 and one corrective prompt per run on average (Table 2). The misses were the familiar classes, Docker in three runs and drag and drop in two. The pattern matches the full prompt at xHigh, where 4 of 6 were first try perfect and both misses included drag and drop. On Opus 4.7, no design prompted cell, at either effort level and with either prompt form, paired the design treatment with the first try reliability of its base counterpart. The one exception in the dataset is the Opus 4.6 abridged cell, where first try reliability was, if anything, higher than base (3 of 6 against 1 of 7). On the newer generation the design treatment is real but consistently priced in reliability as well as tokens, and raising effort does not waive the price.

**Table 7.** Design prompt ablation, no testing tool. The abridged prompt is the one paragraph paraphrase reproduced in the methods. The visual rating column gives the number of runs at each holistic rating (3, 4 and 5; no run in these cells was rated 1 or 2); the rating was not blinded to condition and is descriptive only.

| Cell (no tool) | Runs | Visual rating (runs at 3 / 4 / 5) | Functional score (mean) | First-try perfect | Median cost (USD) |
|---|---|---|---|---|---|
| Opus 4.6 High, base | 7 | 7 / 0 / 0 | 40.6 | 1 of 7 | $2.74 |
| Opus 4.6 High, full design prompt | 6 | 0 / 5 / 1 | 40.5 | 0 of 6 | $3.12 |
| Opus 4.6 High, abridged prompt | 6 | 0 / 6 / 0 | 41.2 | 3 of 6 | $2.77 |
| Opus 4.7 High, base | 6 | 6 / 0 / 0 | 41.0 | 2 of 6 | $3.06 |
| Opus 4.7 High, full design prompt | 6 | 0 / 1 / 5 | 40.5 | 0 of 6 | $4.28 |
| Opus 4.7 High, abridged prompt | 6 | 0 / 0 / 6 | 40.8 | 1 of 6 | $4.14 |
| Opus 4.7 xHigh, base | 6 | 6 / 0 / 0 | 42.0 | 6 of 6 | $3.33 |

| Cell (no tool) | Runs | Visual rating (runs at 3 / 4 / 5) | Functional score (mean) | First-try perfect | Median cost (USD) |
|---|---|---|---|---|---|
| Opus 4.7 xHigh, full design prompt | 6 | 0 / 1 / 5 | 41.5 | 4 of 6 | $5.17 |
| Opus 4.7 xHigh, abridged prompt | 6 | 0 / 0 / 6 | 41.0 | 1 of 6 | $4.73 |

### Within-configuration variability is real but effort-sensitive

Identical configurations did not produce identical software. Across the seven repeated Opus 4.6 runs at High effort, the implementations differed in size and structure as well as styling: the source spanned 10 to 17 files and about 790 to 1320 lines, and the volume of hand-written CSS ranged from 41 to 626 lines, a roughly fifteen-fold spread that was the most variable dimension. Functional scores stayed in a narrow 38 to 42 band, and none of these structural measures tracked the score (Table 8). Raising effort to xHigh, available on Opus 4.7, compressed the functional scatter to a near uniform 42. This is the same run to run instability that others have reported for LLM code generation at fixed settings (Atil et al., 2024; Ouyang et al., 2025), seen here within one task. The practical reading is that a large part of run-to-run variability is a symptom of insufficient deliberation, and that any single agent output should be treated as one sample from a wide distribution rather than as a representative result.

**Table 8.** Within-configuration variability. The seven repeated Claude Opus 4.6 runs at High effort in the base configuration, with no tool or design prompt, sorted by hand authored CSS. Source files counts JavaScript and TypeScript modules. Total source lines and CSS lines count their respective files. All exclude dependencies. The implementations vary in size, structure, and styling, CSS most of all at about fifteen-fold, while functional scores stay between 38 and 42 with no clear relationship to the structural measures.

| Run | Functional score (/42) | Source files | Total source lines | CSS lines |
|---|---|---|---|---|
| 1 | 41 | 17 | 890 | 41 |
| 2 | 41 | 10 | 823 | 53 |
| 3 | 40 | 13 | 1319 | 223 |
| 4 | 41 | 12 | 792 | 466 |
| 5 | 41 | 16 | 869 | 521 |
| 6 | 42 | 11 | 919 | 543 |
| 7 | 38 | 12 | 875 | 626 |

The same measurement on Opus 4.7 shows that this structural spread, like the failure modes, is generation sensitive (Table 9). Across the six repeated Opus 4.7 base runs at High effort the hand-written CSS spans 273 to 395 lines, roughly 1.4-fold against the fifteen-fold spread on Opus 4.6, with 14 to 19 source files and about 1050 to 1490 total source lines, even though the functional scores still scatter from 39 to 42. At xHigh the functional scores are a uniform 42 while the structural spread stays modest, with CSS between 174 and 468 lines. The newer generation therefore converged structurally on its own, while its functional scatter still required the higher effort setting, and in both generations structure and score vary independently, so neither is a proxy for the other.

**Table 9.** Within-configuration variability on Opus 4.7. The six repeated base runs at each effort level, with no tool or design prompt, sorted by hand authored CSS within each effort block. Counting rules match Table 8.

| Run | Effort | Functional score (/42) | Source files | Total source lines | CSS lines |
|---|---|---|---|---|---|
| 1 | High | 41 | 17 | 1201 | 273 |
| 2 | High | 41 | 14 | 1046 | 338 |
| 3 | High | 42 | 19 | 1491 | 355 |
| 4 | High | 39 | 18 | 1388 | 378 |
| 5 | High | 42 | 16 | 1320 | 391 |
| 6 | High | 41 | 16 | 1166 | 395 |
| 1 | xHigh | 42 | 18 | 1048 | 174 |
| 2 | xHigh | 42 | 15 | 1062 | 241 |
| 3 | xHigh | 42 | 17 | 1356 | 281 |
| 4 | xHigh | 42 | 22 | 1396 | 299 |
| 5 | xHigh | 42 | 17 | 1496 | 365 |
| 6 | xHigh | 42 | 18 | 1460 | 468 |

This run-to-run spread is also harder to suppress than it once was, because direct control over sampling randomness has narrowed on current frontier models. Anthropic’s API reference now lists temperature, top_p and top_k as deprecated: models released after Claude Opus 4.6, which include the Opus 4.7 used for the effort contrast, accept only the default temperature of 1 and reject any other value with an error (Anthropic, 2026). The Claude Code harness used here exposes no temperature setting for any model, so the 4.6 generation runs also ran at the default. The direction is industry wide: Google defaults Gemini 3 to a temperature of 1 and advises against lowering it for reasoning tasks (Google, 2026), and OpenAI’s reasoning models, the o-series and the GPT-5 series, do not accept the temperature parameter at all (Microsoft, 2026). Only the local model, Qwen, left temperature fully under the operator’s control. The control a practitioner would normally reach for to make repeated runs more deterministic is therefore unavailable or discouraged on these models, so the run-to-run spread reported here cannot simply be dialed down, and raising reasoning effort, the lever that did compress it, is the one that remains.

## Discussion

The unifying lesson is to match the added resource to the failure mode. In this task the dominant failures were build and environment problems, such as a native SQLite module (better-sqlite3) that does not compile in a minimal container and a breaking change in a web framework (Express 5) that removed a wildcard route. The criterion level profile makes this concrete, with the two environment criteria accounting for over half of all first try failures (Table 3). A tool aimed specifically at these faults, for example a container build or smoke test, could in principle have caught them, so the point is not that no tool could help. It is that the tool added here, a visual automation and testing tool that exercises the running front end, was not matched to where the failures occurred, and the token decomposition shows its cost premium went to context re-reading rather than to any additional engineering. In practice the matching is the hard part, because it requires anticipating which failure mode will arise and adding the corresponding tool, and each added tool adds cost, a trade off that recent work on agent efficiency documents more broadly (Kapoor et al., 2024; Fan et al., 2025; Zhang et al., 2026). Raising reasoning effort, by contrast, improved reliability across failure modes without that guesswork in the base and tool conditions, which is why the same budget spent on effort would have bought the reliability directly. The ablation shows the one boundary of that lever, since a design prompt consumed part or all of the reliability that effort otherwise buys.

### Post repair totals understate differences between models

The criterion level results carry a methodological implication for agent evaluation, and it lines up with a growing criticism of summary scores. Under this rubric, as in most graded evaluations that allow repair, a model that fails and then fixes a criterion keeps most of the credit, so aggregate scores mostly measure what a model can eventually reach rather than what it delivers unaided. Others have raised related concerns: pass@k is often computed in ways that overstate reliability (Jiang et al., 2026), capability and reliability deserve separate measurement in long running agents (Khanal, Tao & Zhou, 2026), and self-repair delivers smaller gains than headline scores suggest (Olausson et al., 2024). What this dataset adds is matched, criterion level evidence of how much a repair tolerant total can hide on a task no model had seen: family means less than a point apart concealed a Docker first try pass rate that moved from 19 percent to 75 percent between frontier generations (a difference of 56 percentage points, 95 percent confidence interval 28 to 73; Table S2), and a failure mode that moved from container builds to package installation rather than disappearing. For teams choosing a model or an effort setting, the deliverable that matters is usually the unaided first attempt, and the number of corrective interventions is a direct human cost. Benchmark reports would be more informative if they published first try reliability and repair burden alongside post repair totals, as others have also urged for reliability and cost (Kapoor et al., 2024; Jiang et al., 2026). In this dataset those axes separate configurations that totals cannot.

### Capability for polish is present but must be invoked

The design prompt result has a counterintuitive shape. Forty-five unprompted runs across three frontier models and two effort levels, and not one imported a display typeface (a named web font such as Inter or Outfit, loaded from Google Fonts), animated an element with a keyframe animation, or drew a CSS gradient, and none defined more than 17 CSS custom properties, the named colour, spacing and type values that a designed interface reuses as design tokens. A one paragraph ask produced all of these in every one of forty prompted runs, each of which imported a display typeface, animated at least one element, drew at least one gradient and defined at least 22 custom properties, even though the prompt contained no detailed visual instructions and asked only, in general terms, for a polished result. The capacity to produce a designed interface is therefore present in the model and does not need to be taught, but it is not expressed unless it is explicitly invoked. Reasoning effort did not invoke it. A short, general design instruction did. The ablation sharpens the point: a one paragraph paraphrase written for this study reproduced the entire treatment on both model generations, so the invocation needs neither length nor emphasis, only the ask itself. That equivalence is itself worth noting, because prompt wording usually matters a great deal: formatting changes alone can move scores by tens of points (Sclar et al., 2024) and rephrasing can change which system looks best (Mizrahi et al., 2024). Here the effect followed the meaning of the ask rather than its wording, which fits evidence that persona style framing and emphasis add little (Zheng et al., 2024). Practitioners who want visual quality should ask for it directly, in as little as a paragraph of project instructions, rather than expecting it to emerge from a larger model or a higher effort setting, and should budget for the small functional risk the more ambitious interface carries in its most interaction heavy features, a risk that persisted at the higher effort setting in the ablation.

### Hidden prompts injected by agents and serving backends

During this work, the Antigravity agent was used, and it was found to inject a system prompt that the user does not see and does not control. Claude Code likewise injects a system prompt that the user does not see. Claude Code's prompt was recovered and inspected for this study, and it contains no visual or design directive, so the base condition carried no competing instruction about appearance. It does, however, instruct the agent to avoid elaboration beyond the task and to exercise interface changes in a browser before reporting completion, the latter overlapping with what the testing tool was meant to supply. This is relevant beyond the present study. When a serving backend or an agent adds its own hidden prompt, the output reflects both the user instruction and the injected text, and a researcher comparing two models through two different backends may be comparing two different hidden prompts as much as two models. The same mechanism is a known security and governance concern, because untrusted injected instructions can change model behavior (OWASP Foundation, 2025). A reasonable recommendation is that anyone comparing models or auditing outputs first establish what hidden prompt, if any, the serving path adds, and report it. The design prompt contrast in this study is, in effect, a controlled version of this phenomenon, isolating the effect of one such injected prompt.

## Limitations

**One task.** The study uses a single task, one conventional full stack web application built from one specification. That trades breadth for depth: ninety runs of one specification buy a level of repetition that multitask benchmarks cannot offer, at the price of not knowing how far the findings travel. The price is real, since skill at one kind of code generation does not necessarily predict skill at another (Du et al., 2023), and improvements tuned to one evaluation setting can transfer weakly to others (Shibaev et al., 2026). Nothing here supports claims about other domains, languages, or task shapes. The findings are best read as a demonstration that these mechanisms exist, with sizes specific to this task.

**One evaluator.** All functional verification was performed by the author, who knew each run's configuration. Most criteria come down to observable outcomes, a build succeeds or fails, data persists or does not, and every run's artifacts and rubric files are archived, so the scoring can be audited. Still, a single evaluator who knows the configuration can lean borderline calls, and describing failure symptoms in corrective prompts involves discretion. Session transcripts were not retained, so the corrective prompts themselves cannot be audited, only the failure each addressed and its outcome, which the rubric files record. Verification also ran on the native Windows host while agents worked in their own environments, so environment dependent faults were scored as failures of the relevant criterion. That is a scoring convention, stated in the methods, as much as a property of the software.

**The holistic visual rating.** The holistic ratings were produced by a single large language model, Claude Opus 4.8 accessed through the Claude desktop application, reviewed by the author, and not produced blind: the model worked from run folders whose names encode the condition, generated its own descriptive scale anchors rather than being given them, and rated the dataset across more than one pass as the sample grew. Model raters also carry known biases (Zheng et al., 2023), agree only partly with human perception of interfaces (Luera et al., 2025), and can favor work produced by their own model family (Panickssery, Bowman & Feng, 2024), which is relevant here because the rater shares a vendor and a model line with most of the interfaces it rated, although it is not itself among the graded models. For these reasons no conclusion rests on the rating. The design prompt result is carried by source-based measures that need no judge (Table 6), and the rating is reported only as a secondary observation whose agreement with those measures, including within the prompted runs, is itself reported. The source-based measures have their own limit: they capture the amount and kind of design work shipped, not whether it looks good, and the correlation between them and the rating is inflated by the between-condition gap, so the within-prompted correlation of 0.59 is the informative figure. A blinded human rating would be needed to make any claim about appearance as such, and is left to future work.

**Ceiling, small cells, and after the fact analyses.** Frontier scores sit near the 42 point ceiling, which limits headroom for detecting improvement, although the null result for the tool is not just a ceiling effect, because at High effort, where scores still scatter between 39 and 42, the tool also failed to help while effort did. Cells of six to seven runs can only reveal large effects, as quantified in the Analysis section. A null result here therefore means that no effect large enough to matter in practice was found, roughly two points or more on the functional score; it does not mean that no effect exists, since a smaller one could be present and go undetected. The xHigh setting exists only on Opus 4.7, so the effort contrast is within a single model family. Conditions were chosen rather than randomized, and the criterion level and token analyses are after the fact examinations of an existing dataset rather than preregistered hypotheses, so the p values and confidence intervals should be read as descriptive. The abridged prompt ablation runs were conducted about a week after the main batches, under the same harness version, so batch timing is a small residual confound for their functional comparison. Session costs are list price totals that vary with the debugging path taken, and the token decomposition covers the 84 runs whose rubrics preserve the per model token breakdown in the end of session report.

**One vendor.** Eighty three of the 90 runs use one vendor’s models under that vendor’s harness, with 85 using its models and 85 its harness, the effort contrast lies within that vendor’s newest family, and the author’s employer is a customer of that vendor, as disclosed in the competing interests statement. No vendor had any role in the study. The specification was also drafted with that vendor’s coding agent, so the task is phrased in the idiom that agent produces, and any advantage this gave the Claude models over the Qwen and Gemini runs cannot be separated from the capability gap reported here. Beyond the disclosure, the concentration itself limits the findings: the generational claims here are claims about one vendor’s release line, and the tool and prompt results are demonstrated on that vendor’s stack. Repeating the study on other vendors’ agent stacks is the natural next step, and the archived task and instrument make that cheap to do.

**Models change over time.** The hosted models studied here are services that change over time, so these exact runs cannot be guaranteed to reproduce even at pinned harness versions. Attempts to reproduce LLM based software engineering studies mostly fail for this reason (Angermeir et al., 2025), on top of ordinary run to run variation (Ouyang et al., 2025). The mitigations are the ones current guidelines recommend (Baltes et al., 2025): a harness held at one version (Claude Code CLI 2.1.132), archived per run rubric files carrying the harness’s end of session usage report, artifacts, and screenshots, and a fully specified task and scoring instrument; full session transcripts were not retained.

These limitations argue for treating the size of each effect as approximate while treating the direction and ordering of effects, which are large and consistent, as robust.

## Conclusions

Across 90 matched and repeated runs of one specification, the choice of model dominated quality, reasoning effort bought first try reliability cheaply, a testing tool bought cost without reliability, and a short design prompt changed what was built, measurably and categorically, without changing function, with a one paragraph version of the prompt producing the same treatment. At the criterion level, post repair totals concealed large differences in first try reliability between frontier generations and a failure mode that shifted rather than disappeared across them, which argues for reporting first try reliability and repair burden alongside final scores. The practical rule that follows is simple. Spend on capability and effort to get reliable software, ask explicitly for visual quality when it matters, and do not expect a verification tool to fix failures it cannot see.

## Acknowledgements

The full design-oriented system prompt studied in the ablation originates with Google's Antigravity agent, whose authors are acknowledged as its source. That prompt is third party material and is not redistributed. It can be recovered from the agent as described in Materials and Methods, and the one paragraph paraphrase written for this study is reproduced there in full.

## Additional information

### Data availability

The full dataset, including every run's source artifacts, screenshots, specification and per run rubric file, the scoring instrument, the analysis code that recomputes every number in this paper and regenerates its tables and figures, the per run visual ratings with the anchors as applied, the per run design treatment measures, the abridged design directive as the runs received it, and an interactive report, is archived on Zenodo under the version specific DOI 10.5281/zenodo.22286311 (Mehta, 2026) and is also available in the project repository on GitHub (https://github.com/achintmehta/retrospective-board-eval). The concept DOI 10.5281/zenodo.20789639 resolves to the latest version. Neither harness's injected system prompt is redistributed, because the licensing terms for their internal prompt text are unclear: the third party design oriented prompt's effect is described in this paper and its abridged paraphrase is reproduced in the methods and archived, and the Claude Code prompt discussed in the discussion is characterized but not reproduced. Both are held local to the source runs and excluded from the archive.

### Supplemental information

Tables S1 and S2 are reproduced in the Appendix. Table S3 and Articles S1 to S3 are included in the Zenodo record cited under Data availability (Mehta, 2026), in the supplements folder of the version specific DOI.

Table S3. Design treatment measures per run: the directive-named features and styling-effort measures counted from each run's shipped source by count_design_markers.py, with the run's condition and holistic visual rating.

Article S1. Derivation of every number reported in this paper from the archived per run rubric files, including the counts behind each Fisher exact test and confidence interval.

Article S2. Completed reporting checklist for empirical studies involving large language models (Baltes et al., 2025).

Article S3. Visual rating instrument: how the holistic ratings were produced, the scale anchors as applied, the distribution over all 90 runs, and per run notes.

## References

Aleithan R, Xue H, Mohajer MM, Nnorom E, Uddin G, Wang S. 2024. SWE-Bench+: enhanced coding benchmark for LLMs. arXiv preprint. DOI 10.48550/arXiv.2410.06992.

Angermeir F, Amougou M, Kreitz M, Bauer A, Linhuber M, Fucci D, et al. 2025. Reflections on the reproducibility of commercial LLM performance in empirical software engineering studies. arXiv preprint. DOI 10.48550/arXiv.2510.25506.

Anthropic. 2026. Create a Message. Claude API reference. Available at https://platform.claude.com/docs/en/api/messages/create (accessed 4 September 2026).

Atil B, Aykent S, Chittams A, Fu L, Passonneau RJ, Radcliffe E, et al. 2024. Non-determinism of “deterministic” LLM settings. arXiv preprint. DOI 10.48550/arXiv.2408.04667.

Austin J, Odena A, Nye M, Bosma M, Michalewski H, Dohan D, et al. 2021. Program synthesis with large language models. arXiv preprint. DOI 10.48550/arXiv.2108.07732.

Bai L, Huang Z, Wang X, Sun J, Mihalcea R, Brynjolfsson E, et al. 2026. How do AI agents spend your money? Analyzing and predicting token consumption in agentic coding tasks. arXiv preprint. DOI 10.48550/arXiv.2604.22750.

Baltes S, Angermeir F, Arora C, Muñoz Barón M, Chen C, et al. 2025. Guidelines for empirical studies in software engineering involving large language models. Empirical Software Engineering, in press. arXiv preprint. DOI 10.48550/arXiv.2508.15503.

Becker J, Rush N, Barnes E, Rein D. 2025. Measuring the impact of early-2025 AI on experienced open-source developer productivity. arXiv preprint. DOI 10.48550/arXiv.2507.09089.

Brown B, Juravsky J, Ehrlich R, Clark R, Le QV, Ré C, Mirhoseini A. 2024. Large language monkeys: scaling inference compute with repeated sampling. arXiv preprint. DOI 10.48550/arXiv.2407.21787.

Chen M, Tworek J, Jun H, Yuan Q, Pinto HPO, Kaplan J, et al. 2021. Evaluating large language models trained on code. arXiv preprint. DOI 10.48550/arXiv.2107.03374.

Chowdhury N, Aung J, Shern CJ, Jaffe O, Sherburn D, Starace G, et al. 2024. Introducing SWE-bench Verified. OpenAI technical report. Available at https://openai.com/index/introducing-swe-bench-verified/ (accessed 27 August 2026).

Du X, Liu M, Wang K, Wang H, Liu J, Chen Y, et al. 2023. ClassEval: a manually-crafted benchmark for evaluating LLMs on class-level code generation. arXiv preprint. DOI 10.48550/arXiv.2308.01861.

Fan Z, Vasilevski K, Lin D, Chen B, Chen Y, Zhong Z, et al. 2025. SWE-Effi: re-evaluating software AI agent system effectiveness under resource constraints. arXiv preprint. DOI 10.48550/arXiv.2509.09853.

Galster M, Mohsenimofidi S, Lulla JL, Abubakar MA, Treude C, Baltes S. 2026. Harness engineering for agentic AI coding tools: an exploratory study. arXiv preprint. DOI 10.48550/arXiv.2602.14690.

Gema AP, Hägele A, Chen R, Arditi A, Goldman-Wetzler J, Fraser-Taliente K, et al. 2025. Inverse scaling in test-time compute. arXiv preprint. DOI 10.48550/arXiv.2507.14417.

Google. 2026. Gemini 3 developer guide (Interactions API). Gemini API documentation. Available at https://ai.google.dev/gemini-api/docs/interactions/gemini-3 (accessed 29 June 2026).

Guo D, Yang D, Zhang H, Song J, Wang P, Zhu Q, et al. 2025. DeepSeek-R1 incentivizes reasoning in LLMs through reinforcement learning. Nature 645(8081):633–638. DOI 10.1038/s41586-025-09422-z.

He H, Miller C, Agarwal S, Kästner C, Vasilescu B. 2026. Speed at the cost of quality: how Cursor AI increases short-term velocity and long-term complexity in open-source projects. In: Proceedings of the 23rd International Conference on Mining Software Repositories (MSR 2026). arXiv preprint. DOI 10.48550/arXiv.2511.04427.

Hou X, Zhao Y, Liu Y, Yang Z, Wang K, Li L, et al. 2024. Large language models for software engineering: a systematic literature review. ACM Transactions on Software Engineering and Methodology. DOI 10.1145/3695988.

Huang J, Chen X, Mishra S, Zheng HS, Yu AW, Song X, Zhou D. 2024. Large language models cannot self-correct reasoning yet. In: International Conference on Learning Representations (ICLR 2024). DOI 10.48550/arXiv.2310.01798.

Jain N, Han K, Gu A, Li W-D, Yan F, Zhang T, et al. 2025. LiveCodeBench: holistic and contamination free evaluation of large language models for code. In: International Conference on Learning Representations (ICLR 2025). DOI 10.48550/arXiv.2403.07974.

Jiang J, Zheng S, Vidra N, Setty S. 2026. Beyond pass@k: measuring reliability and security of agentic code generation. arXiv preprint. DOI 10.48550/arXiv.2608.14711.

Jimenez CE, Yang J, Wettig A, Yao S, Pei K, Press O, Narasimhan K. 2024. SWE-bench: can language models resolve real-world GitHub issues? In: International Conference on Learning Representations (ICLR 2024). DOI 10.48550/arXiv.2310.06770.

Kapoor S, Stroebl B, Siegel ZS, Nadgir N, Narayanan A. 2024. AI agents that matter. arXiv preprint. DOI 10.48550/arXiv.2407.01502.

Khanal A, Tao Y, Zhou J. 2026. Beyond pass@1: a reliability science framework for long-horizon LLM agents. arXiv preprint. DOI 10.48550/arXiv.2603.29231.

Li Y, Choi D, Chung J, Kushman N, Schrittwieser J, Leblond R, et al. 2022. Competition-level code generation with AlphaCode. Science 378(6624):1092–1097. DOI 10.1126/science.abq1158.

Liang S, Garg S, Moghaddam RZ. 2025. The SWE-bench illusion: when state-of-the-art LLMs remember instead of reason. arXiv preprint. DOI 10.48550/arXiv.2506.12286.

Liu J, Xia CS, Wang Y, Zhang L. 2023. Is your code generated by ChatGPT really correct? Rigorous evaluation of large language models for code generation. In: Advances in Neural Information Processing Systems 36 (NeurIPS 2023). DOI 10.48550/arXiv.2305.01210.

Liu J, Wang K, Chen Y, Peng X, Chen Z, Zhang L, Lou Y. 2024. Large language model-based agents for software engineering: a survey. ACM Transactions on Software Engineering and Methodology. DOI 10.1145/3796507.

Luera RA, Rossi R, Dernoncourt F, Basu S, Kim S, Mukherjee S, et al. 2025. MLLM as a UI judge: benchmarking multimodal LLMs for predicting human perception of user interfaces. arXiv preprint. DOI 10.48550/arXiv.2510.08783.

Matton A, Sherborne T, Aumiller D, Tommasone E, Alizadeh M, He J, et al. 2024. On leakage of code generation evaluation datasets. In: Findings of the Association for Computational Linguistics: EMNLP 2024. DOI 10.48550/arXiv.2407.07565.

Mehta A. 2026. Realtime retrospective board: AI model benchmark dataset and evaluation artifacts. Version 2.4.0. Zenodo. DOI 10.5281/zenodo.22286311.

Merrill MA, Shaw AG, Carlini N, Li B, Raj H, et al. 2026. Terminal-Bench: benchmarking agents on hard, realistic tasks in command line interfaces. arXiv preprint. DOI 10.48550/arXiv.2601.11868.

Microsoft. 2026. Azure OpenAI reasoning models: GPT-5 series, o3-mini, o1, o1-mini. Microsoft Foundry documentation. Available at https://learn.microsoft.com/en-us/azure/ai-foundry/openai/how-to/reasoning (accessed 3 September 2026).

Mizrahi M, Kaplan G, Malkin D, Dror R, Shahaf D, Stanovsky G. 2024. State of what art? A call for multi-prompt LLM evaluation. Transactions of the Association for Computational Linguistics 12:933–949. DOI 10.1162/tacl_a_00681.

Muennighoff N, Yang Z, Shi W, Li XL, Fei-Fei L, Hajishirzi H, et al. 2025. s1: simple test-time scaling. In: Proceedings of EMNLP 2025, 20275–20321. DOI 10.18653/v1/2025.emnlp-main.1025.

Olausson TX, Inala JP, Wang C, Gao J, Solar-Lezama A. 2024. Is self-repair a silver bullet for code generation? In: International Conference on Learning Representations (ICLR 2024). DOI 10.48550/arXiv.2306.09896.

Ouyang S, Zhang JM, Harman M, Wang M. 2025. An empirical study of the non-determinism of ChatGPT in code generation. ACM Transactions on Software Engineering and Methodology 34(2). DOI 10.1145/3697010.

OWASP Foundation. 2025. OWASP top 10 for large language model applications: LLM01 prompt injection. Available at https://genai.owasp.org/llmrisk/llm01-prompt-injection/ (accessed 27 June 2026).

Panickssery A, Bowman SR, Feng S. 2024. LLM evaluators recognize and favor their own generations. In: Advances in Neural Information Processing Systems 37 (NeurIPS 2024). DOI 10.48550/arXiv.2404.13076.

Peng S, Kalliamvakou E, Cihon P, Demirer M. 2023. The impact of AI on developer productivity: evidence from GitHub Copilot. arXiv preprint. DOI 10.48550/arXiv.2302.06590.

Ralph P, Ali NB, Baltes S, Bianculli D, Diaz J, et al. 2020. Empirical standards for software engineering research. arXiv preprint. DOI 10.48550/arXiv.2010.03525.

Riddell M, Ni A, Cohan A. 2024. Quantifying contamination in evaluating code generation capabilities of language models. In: Proceedings of the 62nd Annual Meeting of the Association for Computational Linguistics (ACL 2024), 14116–14137. DOI 10.18653/v1/2024.acl-long.761.

Sclar M, Choi Y, Tsvetkov Y, Suhr A. 2024. Quantifying language models' sensitivity to spurious features in prompt design. In: International Conference on Learning Representations (ICLR 2024). DOI 10.48550/arXiv.2310.11324.

Shibaev E, Kudrevskaia V, Galimzyanov T, Evtikhiev M, Terna A, et al. 2026. Don't claim benchmark-oriented optimization improves general coding capability — diverse evaluation is required. In: DL4Code Workshop at ICLR 2026. DOI 10.48550/arXiv.2608.13566.

Si C, Zhang Y, Li R, Yang Z, Liu R, Yang D. 2025. Design2Code: benchmarking multimodal code generation for automated front-end engineering. In: Proceedings of NAACL 2025. DOI 10.48550/arXiv.2403.03163.

Snell C, Lee J, Xu K, Kumar A. 2024. Scaling LLM test-time compute optimally can be more effective than scaling model parameters. arXiv preprint. DOI 10.48550/arXiv.2408.03314.

Song Y, Wang G, Li S, Lin BY. 2025. The good, the bad, and the greedy: evaluation of LLMs should not ignore non-determinism. In: Proceedings of NAACL-HLT 2025, 4195–4206. DOI 10.18653/v1/2025.naacl-long.211.

Vaithilingam P, Zhang T, Glassman EL. 2022. Expectation vs. experience: evaluating the usability of code generation tools powered by large language models. In: CHI Conference on Human Factors in Computing Systems Extended Abstracts (CHI EA 2022). DOI 10.1145/3491101.3519665.

Wang X, Wei J, Schuurmans D, Le Q, Chi E, Narang S, et al. 2023. Self-consistency improves chain of thought reasoning in language models. In: International Conference on Learning Representations (ICLR 2023). DOI 10.48550/arXiv.2203.11171.

Wang X, Li B, Song Y, Xu FF, Tang X, Zhuge M, et al. 2025. OpenHands: an open platform for AI software developers as generalist agents. In: International Conference on Learning Representations (ICLR 2025). DOI 10.48550/arXiv.2407.16741.

Wei J, Wang X, Schuurmans D, Bosma M, Ichter B, Xia F, et al. 2022. Chain-of-thought prompting elicits reasoning in large language models. In: Advances in Neural Information Processing Systems 35 (NeurIPS 2022). DOI 10.48550/arXiv.2201.11903.

Wohlin C, Runeson P, Höst M, Ohlsson MC, Regnell B, Wesslén A. 2012. Experimentation in software engineering. Berlin: Springer. DOI 10.1007/978-3-642-29044-2.

Yang J, Jimenez CE, Wettig A, Lieret K, Yao S, Narasimhan K, Press O. 2024. SWE-agent: agent-computer interfaces enable automated software engineering. In: Advances in Neural Information Processing Systems 37 (NeurIPS 2024). DOI 10.48550/arXiv.2405.15793.

Zhang C, Li Y, Xu C, Liu J, Liu A, Zhou C, et al. 2025. ArtifactsBench: bridging the visual-interactive gap in LLM code generation evaluation. arXiv preprint. DOI 10.48550/arXiv.2507.04952.

Zhang K, Xiong Z, Zhong M, Jiang Z, Yuan Z, Li Z, Lin Y. 2026. Are tools all we need? Unveiling the tool-use tax in LLM agents. arXiv preprint. DOI 10.48550/arXiv.2605.00136.

Zheng L, Chiang W-L, Sheng Y, Zhuang S, Wu Z, Zhuang Y, et al. 2023. Judging LLM-as-a-judge with MT-Bench and Chatbot Arena. In: Advances in Neural Information Processing Systems 36 (NeurIPS 2023), Datasets and Benchmarks Track. DOI 10.48550/arXiv.2306.05685.

Zheng M, Pei J, Logeswaran L, Lee M, Jurgens D. 2024. When “a helpful assistant” is not really helpful: personas in system prompts do not improve performances of large language models. In: Findings of the Association for Computational Linguistics: EMNLP 2024, 15126–15154. DOI 10.18653/v1/2024.findings-emnlp.888.

Zhu Y, Jin T, Pruksachatkun Y, Zhang A, Liu S, Kapoor S, et al. 2025. Establishing best practices for building rigorous agentic benchmarks. arXiv preprint. DOI 10.48550/arXiv.2507.02825.

Zhuo TY, Vu MC, Chim J, Hu H, Yu W, et al. 2025. BigCodeBench: benchmarking code generation with diverse function calls and complex instructions. In: International Conference on Learning Representations (ICLR 2025). DOI 10.48550/arXiv.2406.15877.

# Appendix

Tables S1 and S2 are reproduced here in full. Table S1 lists every run; Table S2 lists the counts behind every inferential comparison in the paper.

Table S1. Per-run data for all 90 runs. Part A: harness, model, reasoning effort, UI testing model or tool as recorded in the rubric, design prompt condition, functional score (out of 42), session cost in US dollars, base-10 logarithm of cost, and holistic visual rating (1 to 5). Part B: the rating of each of the 14 criteria, the total, the first try perfect flag and the number of criteria rated 2.

**Part A.** Per-run configuration, score, cost and visual rating.

| Run folder | Harness | Model | Effort | UI testing model/tool | Design prompt | Score (/42) | Cost (USD) | log10(cost) | Visual rating (1-5) |
|---|---|---|---|---|---|---|---|---|---|
| antigravity_gemini3_flash | Antigravity | Gemini 3.1 Flash | | GPT-OSS 120B | Yes | 38 | | | 3 |
| antigravity_gemini_3.1_pro_high | Antigravity | Gemini 3.1 Pro high | High | Browser subagent | Yes | 40 | | | 2 |
| antigravity_gemini_3.1_pro_low | Antigravity | Gemini 3.1 Pro low | Low | Browser subagent | Yes | 40 | | | 3 |
| antigravity_opus_4.6 | Antigravity | Claude Opus 4.6 | | GPT-OSS 120B | Yes | 40 | | | 3 |
| antigravity_sonnet_4.6 | Antigravity | Claude Sonnet 4.6 | | GPT-OSS 120B | Yes | 42 | | | 4 |
| claude_opus_4.6_high | Claude Code | Claude Opus 4.6 | High | None | No | 41 | $5.61 | 0.749 | 3 |
| claude_opus_4.6_high_abridged_prompt | Claude Code | Claude Opus 4.6 | High | None | Abridged | 42 | $3.45 | 0.538 | 4 |
| claude_opus_4.6_high_abridged_prompt_run_2 | Claude Code | Claude Opus 4.6 | High | None | Abridged | 42 | $2.94 | 0.468 | 4 |
| claude_opus_4.6_high_abridged_prompt_run_3 | Claude Code | Claude Opus 4.6 | High | None | Abridged | 42 | $2.10 | 0.322 | 4 |
| claude_opus_4.6_high_abridged_prompt_run_4 | Claude Code | Claude Opus 4.6 | High | None | Abridged | 41 | $2.31 | 0.364 | 4 |
| claude_opus_4.6_high_abridged_prompt_run_5 | Claude Code | Claude Opus 4.6 | High | None | Abridged | 39 | $3.59 | 0.555 | 4 |
| claude_opus_4.6_high_abridged_prompt_run_6 | Claude Code | Claude Opus 4.6 | High | None | Abridged | 41 | $2.60 | 0.415 | 4 |
| claude_opus_4.6_high_run_1 | Claude Code | Claude Opus 4.6 | High | None | No | 38 | $6.62 | 0.821 | 3 |
| claude_opus_4.6_high_run_2 | Claude Code | Claude Opus 4.6 | High | None | No | 41 | $2.77 | 0.442 | 3 |
| claude_opus_4.6_high_run_3 | Claude Code | Claude Opus 4.6 | High | None | No | 40 | $2.74 | 0.438 | 3 |
| claude_opus_4.6_high_run_4 | Claude Code | Claude Opus 4.6 | High | None | No | 41 | $2.09 | 0.320 | 3 |
| claude_opus_4.6_high_run_5 | Claude Code | Claude Opus 4.6 | High | None | No | 42 | $2.54 | 0.405 | 3 |
| claude_opus_4.6_high_run_6 | Claude Code | Claude Opus 4.6 | High | None | No | 41 | $2.04 | 0.310 | 3 |
| claude_opus_4.6_high_with_antigravity_prompt | Claude Code | Claude Opus 4.6 | High | None | Yes | 41 | $3.12 | 0.494 | 4 |
| claude_opus_4.6_high_with_antigravity_prompt_run_2 | Claude Code | Claude Opus 4.6 | High | None | Yes | 40 | $3.11 | 0.493 | 4 |

| Run folder | Harness | Model | Effort | UI testing model/tool | Design prompt | Score (/42) | Cost (USD) | log10(cost) | Visual rating (1-5) |
|---|---|---|---|---|---|---|---|---|---|
| claude_opus_4.6_high_with_antigravity_prompt_run_3 | Claude Code | Claude Opus 4.6 | High | None | Yes | 40 | $3.51 | 0.545 | 4 |
| claude_opus_4.6_high_with_antigravity_prompt_run_4 | Claude Code | Claude Opus 4.6 | High | None | Yes | 40 | $6.08 | 0.784 | 4 |
| claude_opus_4.6_high_with_antigravity_prompt_run_5 | Claude Code | Claude Opus 4.6 | High | None | Yes | 41 | $2.34 | 0.369 | 5 |
| claude_opus_4.6_high_with_antigravity_prompt_run_6 | Claude Code | Claude Opus 4.6 | High | None | Yes | 41 | $2.72 | 0.435 | 4 |
| claude_opus_4.6_with_playwright_high | Claude Code | Claude Opus 4.6 | High | Playwright | No | 40 | $5.36 | 0.729 | 3 |
| claude_opus_4.6_with_playwright_high_run_2 | Claude Code | Claude Opus 4.6 | High | Playwright | No | 41 | $4.49 | 0.652 | 3 |
| claude_opus_4.6_with_playwright_high_run_3 | Claude Code | Claude Opus 4.6 | High | Playwright | No | 41 | $3.56 | 0.551 | 3 |
| claude_opus_4.6_with_playwright_high_run_4 | Claude Code | Claude Opus 4.6 | High | Playwright | No | 41 | $3.47 | 0.540 | 3 |
| claude_opus_4.6_with_playwright_high_run_5 | Claude Code | Claude Opus 4.6 | High | Playwright | No | 41 | $3.30 | 0.519 | 3 |
| claude_opus_4.6_with_playwright_high_run_6 | Claude Code | Claude Opus 4.6 | High | Playwright | No | 41 | $3.19 | 0.504 | 3 |
| claude_opus_4.6_with_playwright_high_run_7 | Claude Code | Claude Opus 4.6 | High | Playwright | No | 39 | $2.08 | 0.318 | 3 |
| claude_opus_4.7_high | Claude Code | Claude Opus 4.7 | High | None | No | 42 | $3.15 | 0.498 | 3 |
| claude_opus_4.7_high_abridged_prompt | Claude Code | Claude Opus 4.7 | High | None | Abridged | 40 | $3.55 | 0.550 | 5 |
| claude_opus_4.7_high_abridged_prompt_run_2 | Claude Code | Claude Opus 4.7 | High | None | Abridged | 42 | $7.21 | 0.858 | 5 |
| claude_opus_4.7_high_abridged_prompt_run_3 | Claude Code | Claude Opus 4.7 | High | None | Abridged | 41 | $3.96 | 0.598 | 5 |
| claude_opus_4.7_high_abridged_prompt_run_4 | Claude Code | Claude Opus 4.7 | High | None | Abridged | 41 | $4.32 | 0.635 | 5 |
| claude_opus_4.7_high_abridged_prompt_run_5 | Claude Code | Claude Opus 4.7 | High | None | Abridged | 41 | $3.33 | 0.522 | 5 |
| claude_opus_4.7_high_abridged_prompt_run_6 | Claude Code | Claude Opus 4.7 | High | None | Abridged | 40 | $4.66 | 0.668 | 5 |
| claude_opus_4.7_high_run_2 | Claude Code | Claude Opus 4.7 | High | None | No | 41 | $3.13 | 0.496 | 3 |
| claude_opus_4.7_high_run_3 | Claude Code | Claude Opus 4.7 | High | None | No | 41 | $2.22 | 0.346 | 3 |
| claude_opus_4.7_high_run_4 | Claude Code | Claude Opus 4.7 | High | None | No | 41 | $2.99 | 0.476 | 3 |
| claude_opus_4.7_high_run_5 | Claude Code | Claude Opus 4.7 | High | None | No | 39 | $3.96 | 0.598 | 3 |
| claude_opus_4.7_high_run_6 | Claude Code | Claude Opus 4.7 | High | None | No | 42 | $2.79 | 0.446 | 3 |
| claude_opus_4.7_high_with_antigravity_prompt | Claude Code | Claude Opus 4.7 | High | None | Yes | 41 | $3.64 | 0.561 | 5 |
| claude_opus_4.7_high_with_antigravity_prompt_run_2 | Claude Code | Claude Opus 4.7 | High | None | Yes | 41 | $4.08 | 0.611 | 5 |
| claude_opus_4.7_high_with_antigravity_prompt_run_3 | Claude Code | Claude Opus 4.7 | High | None | Yes | 41 | $4.48 | 0.651 | 5 |
| claude_opus_4.7_high_with_antigravity_prompt_run_4 | Claude Code | Claude Opus 4.7 | High | None | Yes | 39 | $4.71 | 0.673 | 4 |
| claude_opus_4.7_high_with_antigravity_prompt_run_5 | Claude Code | Claude Opus 4.7 | High | None | Yes | 40 | $6.57 | 0.818 | 5 |
| claude_opus_4.7_high_with_antigravity_prompt_run_6 | Claude Code | Claude Opus 4.7 | High | None | Yes | 41 | $3.15 | 0.498 | 5 |
| claude_opus_4.7_with_playwright_high | Claude Code | Claude Opus 4.7 | High | Playwright | No | 42 | $4.01 | 0.603 | 3 |
| claude_opus_4.7_with_playwright_high_run_2 | Claude Code | Claude Opus 4.7 | High | Playwright | No | 42 | $3.26 | 0.513 | 3 |
| claude_opus_4.7_with_playwright_high_run_3 | Claude Code | Claude Opus 4.7 | High | Playwright | No | 42 | $4.47 | 0.650 | 3 |

| Run folder | Harness | Model | Effort | UI testing model/tool | Design prompt | Score (/42) | Cost (USD) | log10(cost) | Visual rating (1-5) |
|---|---|---|---|---|---|---|---|---|---|
| claude_opus_4.7_with_playwright_high_run_4 | Claude Code | Claude Opus 4.7 | High | Playwright | No | 41 | $4.22 | 0.625 | 3 |
| claude_opus_4.7_with_playwright_high_run_5 | Claude Code | Claude Opus 4.7 | High | Playwright | No | 41 | $5.03 | 0.702 | 3 |
| claude_opus_4.7_with_playwright_high_run_6 | Claude Code | Claude Opus 4.7 | High | Playwright | No | 41 | $4.88 | 0.688 | 3 |
| claude_opus_4.7_with_playwright_xhigh | Claude Code | Claude Opus 4.7 | xHigh | Playwright | No | 42 | $5.57 | 0.746 | 3 |
| claude_opus_4.7_with_playwright_xhigh_run_2 | Claude Code | Claude Opus 4.7 | xHigh | Playwright | No | 42 | $6.16 | 0.790 | 3 |
| claude_opus_4.7_with_playwright_xhigh_run_3 | Claude Code | Claude Opus 4.7 | xHigh | Playwright | No | 42 | $4.51 | 0.654 | 3 |
| claude_opus_4.7_with_playwright_xhigh_run_4 | Claude Code | Claude Opus 4.7 | xHigh | Playwright | No | 42 | $7.63 | 0.883 | 3 |
| claude_opus_4.7_with_playwright_xhigh_run_5 | Claude Code | Claude Opus 4.7 | xHigh | Playwright | No | 42 | $4.56 | 0.659 | 3 |
| claude_opus_4.7_with_playwright_xhigh_run_6 | Claude Code | Claude Opus 4.7 | xHigh | Playwright | No | 42 | $5.61 | 0.749 | 3 |
| claude_opus_4.7_xhigh | Claude Code | Claude Opus 4.7 | xHigh | None | No | 42 | $3.31 | 0.520 | 3 |
| claude_opus_4.7_xhigh_abridged_prompt | Claude Code | Claude Opus 4.7 | xHigh | None | Abridged | 42 | $4.83 | 0.684 | 5 |
| claude_opus_4.7_xhigh_abridged_prompt_run_2 | Claude Code | Claude Opus 4.7 | xHigh | None | Abridged | 41 | $5.98 | 0.777 | 5 |
| claude_opus_4.7_xhigh_abridged_prompt_run_3 | Claude Code | Claude Opus 4.7 | xHigh | None | Abridged | 41 | $4.89 | 0.689 | 5 |
| claude_opus_4.7_xhigh_abridged_prompt_run_4 | Claude Code | Claude Opus 4.7 | xHigh | None | Abridged | 41 | $3.15 | 0.498 | 5 |
| claude_opus_4.7_xhigh_abridged_prompt_run_5 | Claude Code | Claude Opus 4.7 | xHigh | None | Abridged | 40 | $4.64 | 0.667 | 5 |
| claude_opus_4.7_xhigh_abridged_prompt_run_6 | Claude Code | Claude Opus 4.7 | xHigh | None | Abridged | 41 | $4.60 | 0.663 | 5 |
| claude_opus_4.7_xhigh_run_2 | Claude Code | Claude Opus 4.7 | xHigh | None | No | 42 | $2.54 | 0.405 | 3 |
| claude_opus_4.7_xhigh_run_3 | Claude Code | Claude Opus 4.7 | xHigh | None | No | 42 | $4.12 | 0.615 | 3 |
| claude_opus_4.7_xhigh_run_4 | Claude Code | Claude Opus 4.7 | xHigh | None | No | 42 | $4.22 | 0.625 | 3 |
| claude_opus_4.7_xhigh_run_5 | Claude Code | Claude Opus 4.7 | xHigh | None | No | 42 | $3.36 | 0.526 | 3 |
| claude_opus_4.7_xhigh_run_6 | Claude Code | Claude Opus 4.7 | xHigh | None | No | 42 | $2.77 | 0.442 | 3 |
| claude_opus_4.7_xhigh_with_antigravity_prompt | Claude Code | Claude Opus 4.7 | xHigh | None | Yes | 42 | $5.59 | 0.747 | 5 |
| claude_opus_4.7_xhigh_with_antigravity_prompt_run_2 | Claude Code | Claude Opus 4.7 | xHigh | None | Yes | 42 | $4.77 | 0.679 | 5 |
| claude_opus_4.7_xhigh_with_antigravity_prompt_run_3 | Claude Code | Claude Opus 4.7 | xHigh | None | Yes | 40 | $7.25 | 0.860 | 5 |
| claude_opus_4.7_xhigh_with_antigravity_prompt_run_4 | Claude Code | Claude Opus 4.7 | xHigh | None | Yes | 42 | $4.80 | 0.681 | 5 |
| claude_opus_4.7_xhigh_with_antigravity_prompt_run_5 | Claude Code | Claude Opus 4.7 | xHigh | None | Yes | 42 | $4.99 | 0.698 | 4 |
| claude_opus_4.7_xhigh_with_antigravity_prompt_run_6 | Claude Code | Claude Opus 4.7 | xHigh | None | Yes | 41 | $5.34 | 0.728 | 5 |
| claude_qwen_3.6_high_with_playwright | Claude Code | Qwen 3.6 | High | Playwright | No | 37 | $41.41 | 1.617 | 3 |
| claude_qwen_coder_next_high_with_playwright | Claude Code | Qwen Coder Next | High | Playwright | No | 24 | $178.88 | 2.253 | 3 |

| Run folder | Harness | Model | Effort | UI testing model/tool | Design prompt | Score (/42) | Cost (USD) | log10(cost) | Visual rating (1-5) |
|---|---|---|---|---|---|---|---|---|---|
| claude_sonnet_4.6_high | Claude Code | Claude Sonnet 4.6 | High | None | No | 39 | $1.58 | 0.199 | 3 |
| claude_sonnet_4.6_high_run_2 | Claude Code | Claude Sonnet 4.6 | High | None | No | 41 | $1.40 | 0.146 | 3 |
| claude_sonnet_4.6_high_run_3 | Claude Code | Claude Sonnet 4.6 | High | None | No | 42 | $1.08 | 0.033 | 3 |
| claude_sonnet_4.6_high_with_antigravity_prompt | Claude Code | Claude Sonnet 4.6 | High | None | Yes | 41 | $1.69 | 0.228 | 4 |
| claude_sonnet_4.6_high_with_antigravity_prompt_run_2 | Claude Code | Claude Sonnet 4.6 | High | None | Yes | 41 | $2.27 | 0.356 | 4 |
| claude_sonnet_4.6_high_with_antigravity_prompt_run_3 | Claude Code | Claude Sonnet 4.6 | High | None | Yes | 42 | $1.82 | 0.260 | 4 |
| claude_sonnet_4.6_max_with_antigravity_prompt | Claude Code | Claude Sonnet 4.6 | Max | None | Yes | 41 | $2.90 | 0.462 | 5 |
| claude_sonnet_4.6_with_playwright_high | Claude Code | Claude Sonnet 4.6 | High | Playwright | No | 39 | $4.90 | 0.690 | 3 |
| claude_sonnet_4.6_with_playwright_high_run_2 | Claude Code | Claude Sonnet 4.6 | High | Playwright | No | 42 | $2.57 | 0.410 | 3 |

**Part B.** Criterion ratings per run (3 = passed first try, 2 = fixed after one corrective prompt, 1 = unresolved), total, first-try-perfect flag (1 = all 14 criteria rated 3) and number of criteria rated 2. Criteria: 1 Local dev environment; 2 Docker deployment; 3 Home page; 4 Board creation; 5 User identification; 6 Card interaction; 7 Moving cards; 8 Commenting; 9 Realtime: new card; 10 Realtime: card move; 11 Realtime: new comment; 12 Data persistence; 13 Documentation; 14 CSV export.

| Run folder | 1. Local dev | 2. Docker | 3. Home | 4. Board | 5. User id | 6. Cards | 7. Move | 8. Comment | 9. RT card | 10. RT move | 11. RT comment | 12. Persist | 13. Docs | 14. CSV | Score (/42) | First-try perfect | Criteria rated 2 |
|---|---|---|---|---|---|---|---|---|---|---|---|---|---|---|---|---|---|
| antigravity_gemini3_flash | 2 | 3 | 2 | 2 | 3 | 3 | 3 | 3 | 3 | 3 | 3 | 3 | 2 | 3 | 38 | 0 | 4 |
| antigravity_gemini_3.1_pro_high | 3 | 2 | 3 | 3 | 3 | 3 | 3 | 3 | 3 | 3 | 3 | 3 | 2 | 3 | 40 | 0 | 2 |
| antigravity_gemini_3.1_pro_low | 3 | 2 | 3 | 3 | 3 | 3 | 3 | 3 | 3 | 3 | 3 | 3 | 3 | 2 | 40 | 0 | 2 |
| antigravity_opus_4.6 | 3 | 2 | 3 | 3 | 3 | 3 | 2 | 3 | 3 | 3 | 3 | 3 | 3 | 3 | 40 | 0 | 2 |
| antigravity_sonnet_4.6 | 3 | 3 | 3 | 3 | 3 | 3 | 3 | 3 | 3 | 3 | 3 | 3 | 3 | 3 | 42 | 1 | 0 |
| claude_opus_4.6_high | 3 | 2 | 3 | 3 | 3 | 3 | 3 | 3 | 3 | 3 | 3 | 3 | 3 | 3 | 41 | 0 | 1 |
| claude_opus_4.6_high_abridged_prompt | 3 | 3 | 3 | 3 | 3 | 3 | 3 | 3 | 3 | 3 | 3 | 3 | 3 | 3 | 42 | 1 | 0 |
| claude_opus_4.6_high_abridged_prompt_run_2 | 3 | 3 | 3 | 3 | 3 | 3 | 3 | 3 | 3 | 3 | 3 | 3 | 3 | 3 | 42 | 1 | 0 |
| claude_opus_4.6_high_abridged_prompt_run_3 | 3 | 3 | 3 | 3 | 3 | 3 | 3 | 3 | 3 | 3 | 3 | 3 | 3 | 3 | 42 | 1 | 0 |
| claude_opus_4.6_high_abridged_prompt_run_4 | 3 | 2 | 3 | 3 | 3 | 3 | 3 | 3 | 3 | 3 | 3 | 3 | 3 | 3 | 41 | 0 | 1 |
| claude_opus_4.6_high_abridged_prompt_run_5 | 3 | 2 | 3 | 3 | 3 | 2 | 2 | 3 | 3 | 3 | 3 | 3 | 3 | 3 | 39 | 0 | 3 |
| claude_opus_4.6_high_abridged_prompt_run_6 | 3 | 2 | 3 | 3 | 3 | 3 | 3 | 3 | 3 | 3 | 3 | 3 | 3 | 3 | 41 | 0 | 1 |
| claude_opus_4.6_high_run_1 | 3 | 2 | 3 | 3 | 3 | 3 | 1 | 3 | 2 | 3 | 3 | 3 | 3 | 3 | 38 | 0 | 2 |
| claude_opus_4.6_high_run_2 | 3 | 2 | 3 | 3 | 3 | 3 | 3 | 3 | 3 | 3 | 3 | 3 | 3 | 3 | 41 | 0 | 1 |
| claude_opus_4.6_high_run_3 | 2 | 2 | 3 | 3 | 3 | 3 | 3 | 3 | 3 | 3 | 3 | 3 | 3 | 3 | 40 | 0 | 2 |

| Run folder | 1. Local dev | 2. Docker | 3. Home | 4. Board | 5. User id | 6. Cards | 7. Move | 8. Comment | 9. RT card | 10. RT move | 11. RT comment | 12. Persist | 13. Docs | 14. CSV | Score (/42) | First-try perfect | Criteria rated 2 |
|---|---|---|---|---|---|---|---|---|---|---|---|---|---|---|---|---|---|
| claude_opus_4.6_high_run_4 | 3 | 2 | 3 | 3 | 3 | 3 | 3 | 3 | 3 | 3 | 3 | 3 | 3 | 3 | 41 | 0 | 1 |
| claude_opus_4.6_high_run_5 | 3 | 3 | 3 | 3 | 3 | 3 | 3 | 3 | 3 | 3 | 3 | 3 | 3 | 3 | 42 | 1 | 0 |
| claude_opus_4.6_high_run_6 | 3 | 2 | 3 | 3 | 3 | 3 | 3 | 3 | 3 | 3 | 3 | 3 | 3 | 3 | 41 | 0 | 1 |
| claude_opus_4.6_high_with_antigravity_prompt | 3 | 2 | 3 | 3 | 3 | 3 | 3 | 3 | 3 | 3 | 3 | 3 | 3 | 3 | 41 | 0 | 1 |
| claude_opus_4.6_high_with_antigravity_prompt_run_2 | 3 | 2 | 3 | 3 | 3 | 3 | 3 | 2 | 3 | 3 | 3 | 3 | 3 | 3 | 40 | 0 | 2 |
| claude_opus_4.6_high_with_antigravity_prompt_run_3 | 3 | 2 | 3 | 3 | 3 | 3 | 2 | 3 | 3 | 3 | 3 | 3 | 3 | 3 | 40 | 0 | 2 |
| claude_opus_4.6_high_with_antigravity_prompt_run_4 | 3 | 2 | 3 | 3 | 3 | 2 | 3 | 3 | 3 | 3 | 3 | 3 | 3 | 3 | 40 | 0 | 2 |
| claude_opus_4.6_high_with_antigravity_prompt_run_5 | 3 | 2 | 3 | 3 | 3 | 3 | 3 | 3 | 3 | 3 | 3 | 3 | 3 | 3 | 41 | 0 | 1 |
| claude_opus_4.6_high_with_antigravity_prompt_run_6 | 3 | 2 | 3 | 3 | 3 | 3 | 3 | 3 | 3 | 3 | 3 | 3 | 3 | 3 | 41 | 0 | 1 |
| claude_opus_4.6_with_playwright_high | 3 | 2 | 3 | 3 | 3 | 3 | 2 | 3 | 3 | 3 | 3 | 3 | 3 | 3 | 40 | 0 | 2 |
| claude_opus_4.6_with_playwright_high_run_2 | 3 | 2 | 3 | 3 | 3 | 3 | 3 | 3 | 3 | 3 | 3 | 3 | 3 | 3 | 41 | 0 | 1 |
| claude_opus_4.6_with_playwright_high_run_3 | 3 | 2 | 3 | 3 | 3 | 3 | 3 | 3 | 3 | 3 | 3 | 3 | 3 | 3 | 41 | 0 | 1 |
| claude_opus_4.6_with_playwright_high_run_4 | 3 | 2 | 3 | 3 | 3 | 3 | 3 | 3 | 3 | 3 | 3 | 3 | 3 | 3 | 41 | 0 | 1 |
| claude_opus_4.6_with_playwright_high_run_5 | 3 | 3 | 2 | 3 | 3 | 3 | 3 | 3 | 3 | 3 | 3 | 3 | 3 | 3 | 41 | 0 | 1 |
| claude_opus_4.6_with_playwright_high_run_6 | 3 | 2 | 3 | 3 | 3 | 3 | 3 | 3 | 3 | 3 | 3 | 3 | 3 | 3 | 41 | 0 | 1 |
| claude_opus_4.6_with_playwright_high_run_7 | 3 | 2 | 3 | 3 | 3 | 3 | 1 | 3 | 3 | 3 | 3 | 3 | 3 | 3 | 39 | 0 | 1 |
| claude_opus_4.7_high | 3 | 3 | 3 | 3 | 3 | 3 | 3 | 3 | 3 | 3 | 3 | 3 | 3 | 3 | 42 | 1 | 0 |
| claude_opus_4.7_high_abridged_prompt | 3 | 3 | 3 | 3 | 3 | 3 | 2 | 2 | 3 | 3 | 3 | 3 | 3 | 3 | 40 | 0 | 2 |
| claude_opus_4.7_high_abridged_prompt_run_2 | 3 | 3 | 3 | 3 | 3 | 3 | 3 | 3 | 3 | 3 | 3 | 3 | 3 | 3 | 42 | 1 | 0 |
| claude_opus_4.7_high_abridged_prompt_run_3 | 2 | 3 | 3 | 3 | 3 | 3 | 3 | 3 | 3 | 3 | 3 | 3 | 3 | 3 | 41 | 0 | 1 |
| claude_opus_4.7_high_abridged_prompt_run_4 | 3 | 2 | 3 | 3 | 3 | 3 | 3 | 3 | 3 | 3 | 3 | 3 | 3 | 3 | 41 | 0 | 1 |
| claude_opus_4.7_high_abridged_prompt_run_5 | 3 | 3 | 3 | 3 | 3 | 3 | 2 | 3 | 3 | 3 | 3 | 3 | 3 | 3 | 41 | 0 | 1 |
| claude_opus_4.7_high_abridged_prompt_run_6 | 2 | 3 | 3 | 3 | 3 | 3 | 2 | 3 | 3 | 3 | 3 | 3 | 3 | 3 | 40 | 0 | 2 |
| claude_opus_4.7_high_run_2 | 2 | 3 | 3 | 3 | 3 | 3 | 3 | 3 | 3 | 3 | 3 | 3 | 3 | 3 | 41 | 0 | 1 |
| claude_opus_4.7_high_run_3 | 2 | 3 | 3 | 3 | 3 | 3 | 3 | 3 | 3 | 3 | 3 | 3 | 3 | 3 | 41 | 0 | 1 |
| claude_opus_4.7_high_run_4 | 2 | 3 | 3 | 3 | 3 | 3 | 3 | 3 | 3 | 3 | 3 | 3 | 3 | 3 | 41 | 0 | 1 |
| claude_opus_4.7_high_run_5 | 2 | 3 | 3 | 3 | 3 | 3 | 3 | 3 | 3 | 3 | 3 | 1 | 3 | 3 | 39 | 0 | 1 |
| claude_opus_4.7_high_run_6 | 3 | 3 | 3 | 3 | 3 | 3 | 3 | 3 | 3 | 3 | 3 | 3 | 3 | 3 | 42 | 1 | 0 |
| claude_opus_4.7_high_with_antigravity_prompt | 2 | 3 | 3 | 3 | 3 | 3 | 3 | 3 | 3 | 3 | 3 | 3 | 3 | 3 | 41 | 0 | 1 |
| claude_opus_4.7_high_with_antigravity_prompt_run_2 | 3 | 2 | 3 | 3 | 3 | 3 | 3 | 3 | 3 | 3 | 3 | 3 | 3 | 3 | 41 | 0 | 1 |

| Run folder | 1. Local dev | 2. Docker | 3. Home | 4. Board | 5. User id | 6. Cards | 7. Move | 8. Comment | 9. RT card | 10. RT move | 11. RT comment | 12. Persist | 13. Docs | 14. CSV | Score (/42) | First-try perfect | Criteria rated 2 |
|---|---|---|---|---|---|---|---|---|---|---|---|---|---|---|---|---|---|
| claude_opus_4.7_high_with_antigravity_prompt_run_3 | 3 | 2 | 3 | 3 | 3 | 3 | 3 | 3 | 3 | 3 | 3 | 3 | 3 | 3 | 41 | 0 | 1 |
| claude_opus_4.7_high_with_antigravity_prompt_run_4 | 2 | 3 | 3 | 2 | 3 | 3 | 2 | 3 | 3 | 3 | 3 | 3 | 3 | 3 | 39 | 0 | 3 |
| claude_opus_4.7_high_with_antigravity_prompt_run_5 | 2 | 3 | 3 | 2 | 3 | 3 | 3 | 3 | 3 | 3 | 3 | 3 | 3 | 3 | 40 | 0 | 2 |
| claude_opus_4.7_high_with_antigravity_prompt_run_6 | 2 | 3 | 3 | 3 | 3 | 3 | 3 | 3 | 3 | 3 | 3 | 3 | 3 | 3 | 41 | 0 | 1 |
| claude_opus_4.7_with_playwright_high | 3 | 3 | 3 | 3 | 3 | 3 | 3 | 3 | 3 | 3 | 3 | 3 | 3 | 3 | 42 | 1 | 0 |
| claude_opus_4.7_with_playwright_high_run_2 | 3 | 3 | 3 | 3 | 3 | 3 | 3 | 3 | 3 | 3 | 3 | 3 | 3 | 3 | 42 | 1 | 0 |
| claude_opus_4.7_with_playwright_high_run_3 | 3 | 3 | 3 | 3 | 3 | 3 | 3 | 3 | 3 | 3 | 3 | 3 | 3 | 3 | 42 | 1 | 0 |
| claude_opus_4.7_with_playwright_high_run_4 | 3 | 2 | 3 | 3 | 3 | 3 | 3 | 3 | 3 | 3 | 3 | 3 | 3 | 3 | 41 | 0 | 1 |
| claude_opus_4.7_with_playwright_high_run_5 | 3 | 2 | 3 | 3 | 3 | 3 | 3 | 3 | 3 | 3 | 3 | 3 | 3 | 3 | 41 | 0 | 1 |
| claude_opus_4.7_with_playwright_high_run_6 | 3 | 2 | 3 | 3 | 3 | 3 | 3 | 3 | 3 | 3 | 3 | 3 | 3 | 3 | 41 | 0 | 1 |
| claude_opus_4.7_with_playwright_xhigh | 3 | 3 | 3 | 3 | 3 | 3 | 3 | 3 | 3 | 3 | 3 | 3 | 3 | 3 | 42 | 1 | 0 |
| claude_opus_4.7_with_playwright_xhigh_run_2 | 3 | 3 | 3 | 3 | 3 | 3 | 3 | 3 | 3 | 3 | 3 | 3 | 3 | 3 | 42 | 1 | 0 |
| claude_opus_4.7_with_playwright_xhigh_run_3 | 3 | 3 | 3 | 3 | 3 | 3 | 3 | 3 | 3 | 3 | 3 | 3 | 3 | 3 | 42 | 1 | 0 |
| claude_opus_4.7_with_playwright_xhigh_run_4 | 3 | 3 | 3 | 3 | 3 | 3 | 3 | 3 | 3 | 3 | 3 | 3 | 3 | 3 | 42 | 1 | 0 |
| claude_opus_4.7_with_playwright_xhigh_run_5 | 3 | 3 | 3 | 3 | 3 | 3 | 3 | 3 | 3 | 3 | 3 | 3 | 3 | 3 | 42 | 1 | 0 |
| claude_opus_4.7_with_playwright_xhigh_run_6 | 3 | 3 | 3 | 3 | 3 | 3 | 3 | 3 | 3 | 3 | 3 | 3 | 3 | 3 | 42 | 1 | 0 |
| claude_opus_4.7_xhigh | 3 | 3 | 3 | 3 | 3 | 3 | 3 | 3 | 3 | 3 | 3 | 3 | 3 | 3 | 42 | 1 | 0 |
| claude_opus_4.7_xhigh_abridged_prompt | 3 | 3 | 3 | 3 | 3 | 3 | 3 | 3 | 3 | 3 | 3 | 3 | 3 | 3 | 42 | 1 | 0 |
| claude_opus_4.7_xhigh_abridged_prompt_run_2 | 3 | 2 | 3 | 3 | 3 | 3 | 3 | 3 | 3 | 3 | 3 | 3 | 3 | 3 | 41 | 0 | 1 |
| claude_opus_4.7_xhigh_abridged_prompt_run_3 | 3 | 3 | 3 | 3 | 3 | 3 | 2 | 3 | 3 | 3 | 3 | 3 | 3 | 3 | 41 | 0 | 1 |
| claude_opus_4.7_xhigh_abridged_prompt_run_4 | 3 | 3 | 3 | 3 | 3 | 3 | 2 | 3 | 3 | 3 | 3 | 3 | 3 | 3 | 41 | 0 | 1 |
| claude_opus_4.7_xhigh_abridged_prompt_run_5 | 2 | 2 | 3 | 3 | 3 | 3 | 3 | 3 | 3 | 3 | 3 | 3 | 3 | 3 | 40 | 0 | 2 |
| claude_opus_4.7_xhigh_abridged_prompt_run_6 | 3 | 2 | 3 | 3 | 3 | 3 | 3 | 3 | 3 | 3 | 3 | 3 | 3 | 3 | 41 | 0 | 1 |
| claude_opus_4.7_xhigh_run_2 | 3 | 3 | 3 | 3 | 3 | 3 | 3 | 3 | 3 | 3 | 3 | 3 | 3 | 3 | 42 | 1 | 0 |
| claude_opus_4.7_xhigh_run_3 | 3 | 3 | 3 | 3 | 3 | 3 | 3 | 3 | 3 | 3 | 3 | 3 | 3 | 3 | 42 | 1 | 0 |
| claude_opus_4.7_xhigh_run_4 | 3 | 3 | 3 | 3 | 3 | 3 | 3 | 3 | 3 | 3 | 3 | 3 | 3 | 3 | 42 | 1 | 0 |
| claude_opus_4.7_xhigh_run_5 | 3 | 3 | 3 | 3 | 3 | 3 | 3 | 3 | 3 | 3 | 3 | 3 | 3 | 3 | 42 | 1 | 0 |
| claude_opus_4.7_xhigh_run_6 | 3 | 3 | 3 | 3 | 3 | 3 | 3 | 3 | 3 | 3 | 3 | 3 | 3 | 3 | 42 | 1 | 0 |

| Run folder | 1. Local dev | 2. Docker | 3. Home | 4. Board | 5. User id | 6. Cards | 7. Move | 8. Comment | 9. RT card | 10. RT move | 11. RT comment | 12. Persist | 13. Docs | 14. CSV | Score (/42) | First-try perfect | Criteria rated 2 |
|---|---|---|---|---|---|---|---|---|---|---|---|---|---|---|---|---|---|
| claude_opus_4.7_xhigh_with_antigravity_prompt | 3 | 3 | 3 | 3 | 3 | 3 | 3 | 3 | 3 | 3 | 3 | 3 | 3 | 3 | 42 | 1 | 0 |
| claude_opus_4.7_xhigh_with_antigravity_prompt_run_2 | 3 | 3 | 3 | 3 | 3 | 3 | 3 | 3 | 3 | 3 | 3 | 3 | 3 | 3 | 42 | 1 | 0 |
| claude_opus_4.7_xhigh_with_antigravity_prompt_run_3 | 3 | 2 | 3 | 3 | 3 | 3 | 2 | 3 | 3 | 3 | 3 | 3 | 3 | 3 | 40 | 0 | 2 |
| claude_opus_4.7_xhigh_with_antigravity_prompt_run_4 | 3 | 3 | 3 | 3 | 3 | 3 | 3 | 3 | 3 | 3 | 3 | 3 | 3 | 3 | 42 | 1 | 0 |
| claude_opus_4.7_xhigh_with_antigravity_prompt_run_5 | 3 | 3 | 3 | 3 | 3 | 3 | 3 | 3 | 3 | 3 | 3 | 3 | 3 | 3 | 42 | 1 | 0 |
| claude_opus_4.7_xhigh_with_antigravity_prompt_run_6 | 3 | 3 | 3 | 3 | 3 | 3 | 2 | 3 | 3 | 3 | 3 | 3 | 3 | 3 | 41 | 0 | 1 |
| claude_qwen_3.6_high_with_playwright | 3 | 2 | 3 | 2 | 2 | 3 | 2 | 2 | 3 | 3 | 3 | 3 | 3 | 3 | 37 | 0 | 5 |
| claude_qwen_coder_next_high_with_playwright | 2 | 2 | 3 | 2 | 1 | 2 | 1 | 1 | 1 | 1 | 1 | 3 | 3 | 1 | 24 | 0 | 4 |
| claude_sonnet_4.6_high | 3 | 2 | 3 | 3 | 3 | 3 | 3 | 3 | 2 | 2 | 3 | 3 | 3 | 3 | 39 | 0 | 3 |
| claude_sonnet_4.6_high_run_2 | 2 | 3 | 3 | 3 | 3 | 3 | 3 | 3 | 3 | 3 | 3 | 3 | 3 | 3 | 41 | 0 | 1 |
| claude_sonnet_4.6_high_run_3 | 3 | 3 | 3 | 3 | 3 | 3 | 3 | 3 | 3 | 3 | 3 | 3 | 3 | 3 | 42 | 1 | 0 |
| claude_sonnet_4.6_high_with_antigravity_prompt | 3 | 2 | 3 | 3 | 3 | 3 | 3 | 3 | 3 | 3 | 3 | 3 | 3 | 3 | 41 | 0 | 1 |
| claude_sonnet_4.6_high_with_antigravity_prompt_run_2 | 3 | 2 | 3 | 3 | 3 | 3 | 3 | 3 | 3 | 3 | 3 | 3 | 3 | 3 | 41 | 0 | 1 |
| claude_sonnet_4.6_high_with_antigravity_prompt_run_3 | 3 | 3 | 3 | 3 | 3 | 3 | 3 | 3 | 3 | 3 | 3 | 3 | 3 | 3 | 42 | 1 | 0 |
| claude_sonnet_4.6_max_with_antigravity_prompt | 3 | 3 | 3 | 3 | 3 | 3 | 2 | 3 | 3 | 3 | 3 | 3 | 3 | 3 | 41 | 0 | 1 |
| claude_sonnet_4.6_with_playwright_high | 3 | 2 | 3 | 3 | 3 | 3 | 3 | 1 | 3 | 3 | 3 | 3 | 3 | 3 | 39 | 0 | 1 |
| claude_sonnet_4.6_with_playwright_high_run_2 | 3 | 3 | 3 | 3 | 3 | 3 | 3 | 3 | 3 | 3 | 3 | 3 | 3 | 3 | 42 | 1 | 0 |

Table S2. Inferential comparisons. The 2x2 counts, two-sided Fisher exact p value, risk difference and 95 percent Newcombe confidence interval for every inferential comparison reported in the paper.

| Comparison | Group 1 | Group 2 | Fisher exact p | Risk difference | 95% CI |
|---|---|---|---|---|---|
| Run perfect on first try, by reasoning effort | Opus 4.7 xHigh 16/18 (89%) | Opus 4.7 High 5/18 (28%) | 0.00049 | 61.1 pp | 29.4 to 78.4 |

| Comparison | Group 1 | Group 2 | Fisher exact p | Risk difference | 95% CI |
|---|---|---|---|---|---|
| Docker criterion passed first try, by model generation | Opus 4.7 High 18/24 (75%) | Opus 4.6 High 5/26 (19%) | 0.00016 | 55.8 pp | 28.5 to 72.6 |
| Drag-and-drop missed first try, by design prompt | Design prompted 11/40 (28%) | No design prompt 5/45 (11%) | 0.0934 | 16.4 pp | −0.4 to 33.0 |